\documentclass[10pt]{article}

\usepackage{amsmath}
\usepackage{amsfonts}
\usepackage{amssymb}
\usepackage{epsfig}
\usepackage{graphicx}
\usepackage{subfigure}
\usepackage{cite}

\begin{document}

\newcommand{\be}{\begin{equation}}\newcommand{\ee}{\end{equation}}
\newcommand{\bea}{\begin{eqnarray}} \newcommand{\eea}{\end{eqnarray}}
\newcommand{\ba}[1]{\begin{array}{#1}} \newcommand{\ea}{\end{array}}

\numberwithin{equation}{section}

\def\pd{\partial}
\def\a{\alpha}
\def\b{\beta}
\def\m{\mu}
\def\mm{m}
\def\n{\nu}
\def\r{\rho}
\def\s{\sigma}
\def\d{\delta}
\def\wg{\wedge}
\def\eps{\epsilon}
\def\veps{\varepsilon}
\def\nn{\nonumber}
\def\ov{\over }
\def\td{\tilde }
\def\Omm{\mu} 

\rightline{ICCUB-10-104}

\rightline{November, 2010}

\bigskip

\begin{center}

{\Large\bf  
$p$-wave Holographic Superconductors and
\\  five-dimensional gauged Supergravity
}
\bigskip
\bigskip

{\it \large Francesco Aprile \footnotemark[1], Diego Rodriguez-Gomez \footnotemark[2] \footnotemark[3] and Jorge G. Russo \footnotemark[1] \footnotemark[4]}
\bigskip

{\it
1) Institute of Cosmos Sciences and Estructura i Constituents de la Materia\\
Facultat de F{\'\i}sica, Universitat de Barcelona\\
Av. Diagonal 647,  08028 Barcelona, Spain\\
\smallskip
2) Department of Physics, \\
Technion, Haifa, 3200, Israel\\
\smallskip
3) Department of Mathematics and Physics, \\
University of Haifa at Oranim, Tivon, 36006, Israel\\
\smallskip
4) Instituci\'o Catalana de Recerca i Estudis Avan\c cats (ICREA)\\
Pg. Lluis Companys, 23, 08010 Barcelona, Spain\\
}
\bigskip
\bigskip

\end{center}
\bigskip

\begin{abstract}
We explore five-dimensional  ${\cal N}=4$ $SU(2)\times U(1)$ and  ${\cal N}=8$ $SO(6)$ gauged  supergravities
as frameworks for condensed matter applications.
These theories contain charged (dilatonic) black holes and
 2-forms which have non-trivial quantum numbers with respect
 to $U(1)$ subgroups of $SO(6)$. A question of interest is whether they also contain black holes with two-form hair
with the required asymptotic to  give rise to holographic superconductivity.
We first consider the  ${\cal N}=4$ case, which contains a 
 complex two-form potential $A_{\mu\nu}$ which has $U(1)$ charge $\pm 1$.  
We find that a slight generalization, where the two-form potential has an arbitrary charge $q$,
leads to a five-dimensional model that exhibits  second-order superconducting transitions of $p$-wave type where 
the role of order parameter is played by $A_{\mu\nu}$, provided $q \gtrsim 5.6$.
We identify the operator that condenses in the dual CFT, which is closely related to ${\cal N}=4$ Super Yang-Mills 
theory with chemical potentials. 
Similar phase transitions between R-charged black holes and black holes with 2-form hair 
are found in a generalized version of the ${\cal N}=8$  gauged supergravity Lagrangian where the two-forms have charge 
$q\gtrsim  1.8$. 

\end{abstract}

\clearpage

\tableofcontents

\section{Introduction}

The application of AdS/CFT to the study of condensed matter systems
is in rapid evolution. One of the most interesting applications has been the investigation of strongly coupled systems which undergo a superconducting
phase transition below a critical temperature \cite{Gubser:2008px,Hartnoll:2008vx,Gubser:2008zu,Hartnoll:2008kx,Gubser:2008pf}.
 On the field theory side, the onset of superconductivity is characterized by the condensation
of a composite charged operator  for low temperatures $T < T_c$. In the dual 
gravitational description, the superconducting phase transition is represented by a transition from a black hole in anti-de Sitter space to
a new black hole solution with ``hair", which is thermodynamically preferred below the critical temperature $T_c$.

Most of the works have adopted a phenomenological approach, where the gravitational system is constructed \textit{ad hoc} and
the underlying field theory is unknown. Although these scenarios have led to interesting qualitative results (see e.g. 
\cite{Horowitz:2008bn,Franco:2009yz,Franco:2009if,Aprile:2009ai,Aprile:2010yb,Benini:2010pr,Faulkner:2010gj}  and \cite{Herzog:2009xv,Horowitz:2010gk} for reviews and references),
understanding the precise dictionary between gravity and the condensed matter system is obviously important in order to make further progress, in particular, for eventual
applications to real systems. In the AdS/CFT context, having a precise dictionary typically requires a brane construction, so that the field theory undergoing the phase transition can be explicitly constructed while at the same time the dual gravitational background can be found. In other words, a top-down approach where one would start with some compactification of type II string or M-theory possibly with some branes present, and consider the dynamics of excitations around those solutions. 
At the linearized level, an example was first given in \cite{Denef:2009tp} by Kaluza-Klein reduction of $D = 11$ supergravity on a seven-dimensional
Sasaki-Einstein  space. Explicit examples of such compactifications leading to systems that exhibit superconducting phase transitions have then appeared in 
 \cite{Gubser:2009qm,Gauntlett:2009dn,Gauntlett:2009bh,Donos:2010ax} (see   \cite{Bobev:2010ib,Arean:2010wu} for further developments).\footnote{
 A different approach where one has some control over the dual field theory is based on using D-brane probes in  string-theory black brane backgrounds (see e.g. \cite{Ammon:2009fe,Peeters:2009sr}).}
 
In  holographic superconducting models, the spontaneously broken $U(1)$ symmetry is typically 
dual to a {\it global} $U(1)$ symmetry in the boundary theory. Consider, in particular,  the top-down constructions of \cite{Gubser:2009qm,Gauntlett:2009dn,Gauntlett:2009bh,Donos:2010ax}.
 These consistent truncations of  IIB/11d supergravity  are closely related to the near horizon region of branes probing Sasaki-Einstein  cones, which are non-compact conical Calabi-Yau spaces.  D3 branes (M2 branes in the 11d case) probing such cones yield to examples of the AdS/CFT correspondence which generically preserve 4 supercharges. In this case the dual field theory can be explicitly constructed\footnote{The identification of the field theory is more direct in the D3 brane case. 
For SCFT's dual to M2 branes probing $CY_4$ the field-theory description is less clear (see e.g. \cite{Jafferis:2008qz}
for a discussion).
Some examples can be found in \cite{Franco:2008um, Franco:2009sp}.}
using by now standard methods (see \textit{e.g.} \cite{Franco:2005sm}), thus providing the desired microscopic theory. For four unbroken supercharges, the generic R-symmetry of these field theories is precisely $U(1)$. It is this particular $U(1)_R$ the one which is spontaneously broken by the condensation of a scalar
representing a breathing mode. 

As the number of preserved supersymmetries is increased, the R-symmetry of the boundary theory is enhanced. For the maximal rigid SUSY in four dimensions, namely $\mathcal{N}=4$, the corresponding R-symmetry is $SU(4)\sim SO(6)$. Thus, while for $\mathcal{N}=1$ field theories there is just one single --and uniquely fixed by superconformal invariance-- $U(1)_R$ symmetry in the infrared, for $\mathcal{N}=4$ Super Yang-Mills theory there are various $U(1)$ generators in the larger non-abelian R-symmetry which might be spontaneously broken leading to a superconducting phase transition. Motivated by this observation, 
in this paper we investigate holographic phase transitions within the framework of five-dimensional $\mathcal{N}=8$ $SO(6)$    gauged supergravity \cite{Gunaydin:1984qu,Gunaydin:1985cu},
$\mathcal{N}=4$ $SU(2)\times U(1)$   gauged supergravity    \cite{Romans:1985ps} 
and in related models.

$\mathcal{N}=4$ $SU(2)\times U(1)$   gauged supergravity can be derived from a consistent truncation of IIB supergravity \cite{Lu:1999bw} and
also from a consistent truncation of eleven dimensional supergravity \cite{Gauntlett:2007sm}.
$\mathcal{N}=8$ $SO(6)$ five-dimensional 
gauged supergravity is expected to arise from a consistent reduction of IIB supergravity on an $S^5$, although in this case an explicit construction is presently unknown.  
As such, these theories provide an interesting arena to study the dynamics associated with the $U(1)^3\in SO(6)$ R-symmetry of interest. 
These theories  contain complex two-form fields (whose ten-dimensional origin is the complex two-form potential of IIB supergravity) which are charged under the  $U(1)$  gauge groups. 
An exciting possibility is that these complex two-form fields could condense and lead to superconductivity in the dual field theory.
We will show that this does not occur, basically because the charge of the two-form is not sufficient large to trigger an instability.
Indeed, a slight generalization of the supergravity theory, whereby we allow the $U(1)$ charge $q$ of the two-forms to take generic values, contains phases where the black hole develops non-trivial two-form hair, thus breaking $U(1)$ symmetry spontaneously. Interestingly, the order parameter transforms as a vector under $SO(3)$ spatial rotations. Therefore the system represents a $p$-wave superconductor. This is of course expected,
 since in five dimensions a 2-form potential is Hodge dual to a one-form potential. While for the free theory a 2-form is dynamically 
equivalent to a 1-form, this is not the case in  the gauged supergravity.
Models of holographic $p$-wave superconductors have been constructed first in \cite{Gubser:2008wv,Roberts:2008ns} and then different versions have appeared 
(see e.g.  \cite{Ammon:2009fe,Peeters:2009sr,Basu:2009vv,Ammon:2009xh}).

The model has well-known charged black hole solutions, the STU black holes  \cite{Behrndt:1998jd}. 
These are  characterized by the ADM mass and  charges $(Q_1,\, Q_2,\, Q_3)$ under the maximal abelian subgroup of the gauge symmetry, namely $U(1)^3$. From a ten-dimensional standpoint, these solutions correspond to spinning black D3 branes in flat $\mathbb{R}^6$ space, and the three electric charges correspond to the three independent angular momenta in $\mathbb{R}^6$. 

In this paper we give a detailed investigation of the different phase transitions that take place as finite charge densities are turned on for different combinations of such $U(1)$
gauge fields. As the  temperature is gradually decreased, it  will be seen that new phases where two-form hair grows become thermodynamically favored. The transition to these hairy phases is second order. 
Their onset
can be determined by solving the equations of motion of the 2-form in the vicinity of the phase transition. This amounts to studying
the emergence of regular zero-modes in the black hole backgrounds, see \cite{Gubser:2008px}. 
The generalization of the model to arbitrary charge $q$ permits
a study of the system in the probe limit where $q$ is large.
Nevertheless, since we are specially interested in the case of supergravity where $q=1$
 a full analysis including back-reaction will be provided.

The thermodynamics of STU black holes has been widely studied in \cite{Gubser:1998jb, Cvetic:1999ne, Cvetic:1999rb}. 
The present models contain additional degrees of freedom,
charged  2-form fields, which will lead to radically different phase transition dynamics.

The organization of this paper is as follows. In section 2, we shall first investigate  $\mathcal{N}=4$ 
$SU(2)\times U(1)$ five-dimensional gauged supergravity (which can be thought as a subsector of the full  $\mathcal{N}=8$ 5d gauged supergravity). This has the advantage that the Lagrangian contains a single scalar field
(and in this sense it is one of the simplest top-down models that one can study).
There are three  distinct ${\cal N}=4$ $SU(2)\times U(1)$ gauged  supergravity theories, 
depending on the values of the $SU(2)$ and $U(1)$ coupling constants. Here we consider the ${\cal N}=4^+$ version, where both coupling constants have the same sign and the theory contains AdS vacua (in the second theory the coupling constants have opposite signs and in the third theory the $ SU(2)$ coupling constant is taken to zero). After a brief review of the gravity theory,  we shall introduce the relevant black hole solution and review some of its salient thermodynamical features. In particular, we will see that these black holes have a Hawking temperature which cannot be less than some minimal value. 
This property further motivates the search for two-form condensation, as a possible avenue that the system can take to go to lower $T$. 
In section 3 we give the ansatz for the hairy black hole solution representing the condensed phase and identify two conserved charges.
Section  4 is devoted to the numerical analysis of the solutions. We first find the critical temperature as a function of the $U(1)$ charge of the 2-form
field by studying the system in the vicinity of the transition,  and then solve the the full system including back reaction, compute the free energy
and describe the different phases in detail. 
We also discuss the probe limit and compute the conductivity, showing that the condensed phase exhibits  transport properties which are characteristic of superconducting materials.
In Section 5 we discuss some features of the dual field theory.
The close connection of the gravity model with gauged supergravity is used to make a concrete proposal for the dual operator that condenses.
In Section 6, we consider the more general case of STU black holes with  $(Q_1,\, Q_2,\, Q_3)\ne 0$ and discuss 
in detail the ${\cal N}=8$ gauged supergravity setup. 
In particular, we consider the case $Q_1=Q_2=Q_3$ and compute the critical temperature as a function of the $U(1)$ charge of the two-form fields,
and show that condensation requires a minimum charge which is above the values that one finds in supergravity.
Some concluding remarks are given in Section 7.

\section{ Condensed matter from ${\cal N}=4$ $SU(2)\times U(1)$ gauged  \\ supergravity: Basic setup}

The $\mathcal{N}=4^+$ gauged supergravity in five dimensions \cite{Romans:1985ps} has a bosonic sector containing the metric, a scalar, 
a $U(1)$ vector field $B_{\m}$, $SU(2)$ Yang-Mills vector fields $A^a_{\m}$,  and two $2$-forms  $A^{\a}_{\m\n}$, $\a=1,2$,  which transform as a charged doublet under  $U(1)$ transformations. The action is given by
\bea\label{Roman}
I &=& -{1\over 16\pi G_N} \int \bigg[R\ast 1-3X^{-2}\ast dX\wg dX-\frac{1}{2}X^4 \ast F_{(2)}\wg F_{(2)}-\frac{1}{2}X^{-2}\Big(\ast G^a_{(2)}\wg G^a_{(2)}+\ast A^{\a}_{(2)}\wg A^{\a}_{(2)}\Big)\nn\\ \nn \\
&&+\ \frac{L}{2}\eps_{\a\b}A^{\a}_{(2)}\wg dA^{\b}_{(2)}-\frac{1}{2}A^{\a}_{(2)}\wg A^{\a}_{(2)}\wg B_{(1)}-\frac{1}{2}G_{(2)}^a \wg G^{a}_{(2)} \wg B_{(1)}
\nn\\
&&+\ \frac{4}{L^2}(X^2+2X^{-1})\ast 1 \bigg]\ .
\eea
The field strengths are 
\be
G_{(2)}^{a}=dA^{a}_{(1)}+\frac{1}{\sqrt{2}L}\eps^{abc}A^{b}_{(1)}\wg A^{c}_{(1)},\qquad F_{(2)}=dB_{(1)}\ ,
\ee
\be
F_{(3)}\equiv DA_{(2)}=dA_{(2)}-\frac{i}{L}A_{(2)}\wg B_{(1)}\ ,
\ee
where we introduced complex notation
\be
A_{(2)}\equiv A_{(2)}^1+iA_{(2)}^2\ .
\ee
The equations of motion derived from (\ref{Roman}) are \cite{Romans:1985ps,Lu:1999bw},\\
\bea
R_{\m\n}&=& 3X^{-2}\pd_{\m}X\pd_{\n}X-\frac{4}{3L^2}(X^2+2X^{-1})g_{\m\n}+\frac{1}{2}X^{-2}(\bar{A}_{(\m}^{\ \ \r}A_{\n)\r}-\frac{1}{6}g_{\m\n}|A_{(2)}|^2) \label{EqR}
 \\ 
& &+\frac{1}{2}X^4(F_{\m}^{\ \r}F_{\n\r}-\frac{1}{6}g_{\m\n}F_{(2)}^2)+\frac{1}{2}X^{-2}(G_{\m}^{a\ \r}G_{\n\r}^a-\frac{1}{6}g_{\m\n}(G^a_{(2)})^2)\ ,\nn \\
d(X^{-1}\ast dX)&=& \frac{1}{3}X^4\ast F_{(2)}\wg F_{(2)}-\frac{1}{6}X^{-2}(\ast G_{(2)}^a\wg G_{(2)}^a+\ast \bar A_{(2)}\wg A_{(2)})-\frac{4}{3L^2}(X^2-X^{-1})\ast 1\ , \nn \\
d(X^4\ast F_{2})&=&-\frac{1}{2}G_{(2)}^a\wg G_{(2)}^a-\frac{1}{2}\bar{A}_{(2)}\wg A_{(2)}\ ,\nn \\
d(X^{-2}\ast G_{(2)}^a)&=&\frac{\sqrt{2}}{L}X^{-2}\eps^{abc}\ast G_{(2)}^b\wg A_{(1)}^c-G_{(2)}^a\wg F_{(2)}\ , \nn\\
X^{2}\ast F_{(3)}&=&\frac{i}{L}A_{(2)}\ .\label{EqA} \nn
\eea
{}From the above equations we see that the non-abelian gauge fields can be consistently set to zero.
 Thus, for our purposes, we can just consider  the Lagrangian
\bea
\mathcal{L} &=&\sqrt{g}\ \Big[\ 
R-3X^{-2}\pd_{\m}X\pd^{\mu}X-\frac{X^4}{4}F_{\m\n}F^{\m\n}+\frac{4}{L^2}(X^2+2X^{-1})\Big]
\nn\\
&+&
\frac{L}{8i}\eps^{\m\n\r\s\d}\bar A_{\m\n}\pd_{\r}A_{\s\d}-\frac{1}{8}\eps^{\m\n\r\s\d}\bar A_{\m\n} A_{\r\s}
B_{\d}-\frac{\sqrt{g}}{4X^2}\bar A_{\m\n} A^{\m\n}\ .
\label{SRoman}
\eea

\subsection{The charged black hole
}
\def\p{\partial }
\def\s{\sigma }

The charged black hole solution can be obtained as a particular case of the STU black hole carrying three different $U(1)$ charges $Q_1,Q_2,Q_3$ \cite{Behrndt:1998jd}.
In the present case we have $Q_2=Q_3=0$, $Q_1\equiv Q$, and the solution becomes
\be
ds^2= -  {f\over H^{2/3}}\, dt^2 + {H^{1/3} \over f}\, dr^2 + H^{1/3} {r^2\over L^2}\ (dx^2+ dy^2+dz^2)\ ,
\ee
\be
X=H^{1/3}\ ,\qquad B_0 = {Q\sqrt{\mm}\over r_h^2+Q^2}- {Q\sqrt{\mm }\over r^2+Q^2} \ ,
\label{atun}
\ee
$$
H= 1+ {Q^2\over r^2}\ ,\qquad f= {r^2\over L^2} +{Q^2\over L^2} -{\mm\over r^2}\ ,
$$
where $r_h$ is the position of the event horizon located at
\be
r_h^2 +Q^2 ={\mm L^2\over r_h^2}\qquad \rightarrow \qquad r_h^2={1\over 2} \left(\sqrt{Q^4+4\mm L^2 } -Q^2\right)\ .
\label{hori}
\ee
Unlike the Reissner-N\" ordstrom solution, this geometry does not have a Cauchy horizon,  the causal structure
is as in the Schwarzschild black hole. There is  a curvature singularity at $r=0$.

The ADM mass  and entropy density for the black hole solution are given by
\be
{M\over V_3}= {1\over 8\pi G_N L^3} \big(Q^2+{3\over 2}\mm \big)\ ,\qquad 
s= {A_h\over 4 G_N V_3} = {r_h\sqrt{\mm } \over 4G_N L^2} \ ,
\ee
By demanding regularity in the Wick rotated solution, one shows that the Hawking temperature is given by
\be
T=
{\sqrt{Q^4+4\mm L^2} \over 2\pi L^2 \sqrt{r_h^2+Q^2}}\ .
\label{hawk}
\ee
In the limit $\mm\to 0$ with $Q$ fixed, one has $r_h\to 0$ and 
the entropy vanishes. Although the Hawking temperature seems to attain a finite value
\be
T_{m=0} = {Q \over 2\pi L^2 }\ ,
\label{temin}
\ee
the classical thermodynamics is no longer reliable because the solution is singular.
 As  explained below, this limit is never reached in studying the field theory thermodynamics, where the choice of a definite ensemble implies  fixing either chemical potential or charge density (which are given in terms of $\{Q,\, m\}$). In particular, note that  physical electric charge is proportional to $Q\sqrt{m}$, so keeping the charge fixed at $m\to 0$ requires
 $Q\to\infty $.


\subsection{Field theory thermodynamics}

The energy and charge densities of the field theory are given by
\be
\epsilon = {3\over 16 \pi G_N L^3}\, \mm \ ,\qquad \rho = {Q\sqrt{\mm }\over 8\pi G_N L^2}\ .
\ee
{} For the sake of simplicity in the formulas, it is convenient to introduce  new variables  rescaling by a factor $2G_N/(\pi L^3)$ as in \cite{Gubser:2009qt}.
One  obtains
\be\label{TermoQuant}
\hat\epsilon = {3\mm\over 8\pi^2 L^6}\ ,\qquad \hat s={r_h\sqrt{\mm}\over 2\pi L^5}\ ,\qquad \hat \rho ={Q\sqrt{\mm }\over 4\pi^2L^5}\ .
\ee
Then the equation (\ref{hori}) determining the location of the horizon gives the microcanonical equation of state:
\be
\hat\epsilon =     {3 \over 2(2\pi)^{2/3}} \left( \hat s^4 + 4\pi^2 \hat s^2\hat\rho^2\right)^{1/3}\ .
\ee
The temperature and chemical potential are then given by
\bea
\label{trho}
T &=& \left( {\partial \hat \epsilon\over\partial\hat s}\right)_{\hat\rho}= {2^{1/3}(\hat s^3 + 2\pi^2 \hat s \hat \rho^2)\over 
 \pi^{2/3} (\hat s^4 + 4\pi^2\hat s^2\hat\rho^2)^{2/3}}
\ ,
\\
\label{tomega}
\Omm  &=& \left( {\partial \hat \epsilon\over\partial\hat \rho}\right)_{\hat s}={ (2\pi)^{4/3}\hat s^2\hat\rho \over  (\hat s^4 + 4\pi^2\hat s^2\hat\rho^2)^{2/3}}
\ .
\eea
They obey the simple relation,
\be
{T\over \Omm  } = {\hat s\over 2\pi^2 \hat\rho} + {\hat\rho\over \hat s} \ 
\label{simrel}
\ee
i.e.
\be
\hat s_{1,2} =\frac{\pi^2 \hat \rho T}{ \Omm  }\left(  1 \mp  \sqrt{ 1-{2 \Omm ^2\over \pi^2T^2}
   } \right)\ .
  \label{tres} \ee
One can check that
\be
\label{chempot}
\Omm  = {1\over L} \ {Q\sqrt{\mm}\over r_h^2+Q^2}\ .
\ee
This agrees with the identification  derived from the standard rules of AdS/CFT
using the asymptotic of the electromagnetic potential.

%


In this paper we will work at fixed  charge density  fixed $\hat \rho $, which corresponds to specifying 
the canonical ensemble.  
In this case the thermodynamics is dominated by the configuration with minimum Helmholtz free energy, given by
\be
F = \hat \epsilon - T\hat s\ .
\label{helm}
\ee


The specific heat at constant charge density is given by
\be
\label{crho}
\hat C_{\hat\rho} = T\left( {\partial \hat s\over\partial T}\right)_{\hat\rho} = \frac{3 \hat s \left(\hat s^2+2\pi ^2 \hat \rho ^2\right)
   \left(\hat s^2+4 \pi ^2 \hat \rho ^2\right)}{\hat s^4+10 \pi ^2
  \hat \rho ^2 \hat s^2-8\pi ^4 \hat \rho ^4}\ ,
\ee

The expression $T=T(\hat s, \hat \rho)$ given in (\ref{trho}) defines $\hat s(T,\hat\rho)$, 
though an explicit formula requires finding the roots of a six-order polynomial. 
However, the main features can be exhibited by a simple numerical analysis.
Figure 1 is a plot $T$ vs. $\hat s$ at $\hat \rho =1$.  We see that the temperature has a minimal value, which is given by
\be
T_{\rm min}=\frac{\sqrt{3} (1122 \sqrt{33}-5758)^{1/6} \hat \rho ^{1/3}}{4
   {\pi }^{1/3}}\approx 0.88\  \hat \rho^{1/3}\ .
 \label{Tmini}
 \ee
It should be stressed that this minimum temperature is  unrelated to the minimum temperature (\ref{temin}), for the reasons explained earlier (in short, the minimum of a function $T$ of two variables $Q,\ \mm $ changes according to which combination of variables is kept fixed). In particular,  from the expression for the charge density in (\ref{TermoQuant}) one sees that the limit $\mm\rightarrow 0$ with fixed $\hat \rho $ requires that $Q\to\infty $ in such a way that the temperature $T$ goes to infinity.

\begin{figure}[h!]
\centering
\includegraphics[scale=.9]{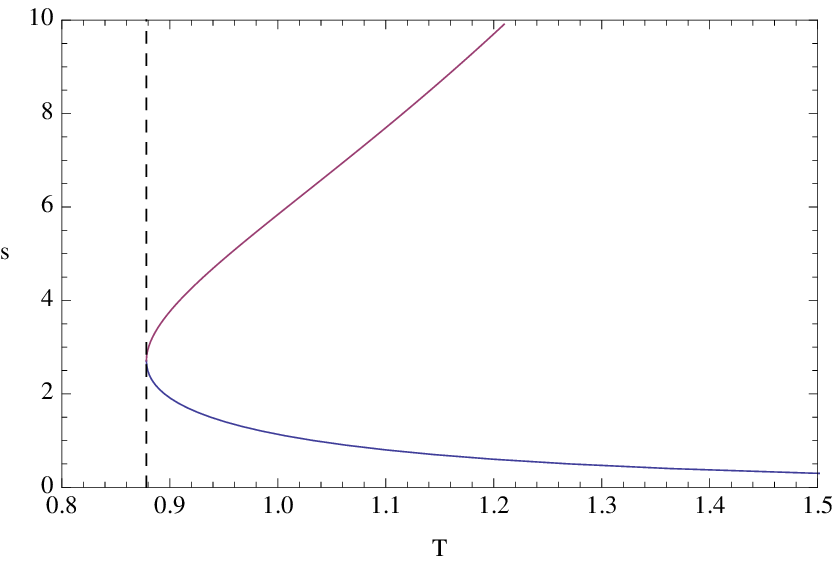} \hspace{.2cm}
\includegraphics[scale=.9]{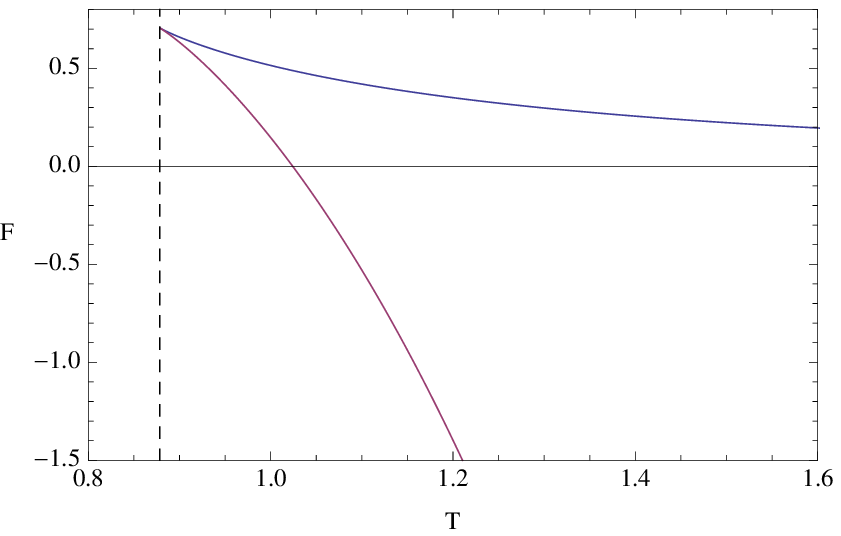} 
\caption{ (Left) Entropy vs. Temperature at fixed  $\hat \rho=1$. (Right)
Free energy as a function of the temperature.  The dashed vertical line  stands for $T_{\rm min}$.
Branch 2 (red) is the upper branch on the Left panel, while branch 1 (blue) is the upper branch on the Right panel.
 }
\label{freerho}
\end{figure}

For $T>T_{\rm min}$ there are two branches, $\hat s_1(T,\hat\rho)$ and $\hat s_2(T,\hat\rho)$, the latter being the one with higher entropy density. The physically relevant branch that has less free energy is $\hat s_2(T,\hat\rho)$, as can be seen from figure \ref{freerho}. This stable branch is  energetically (and entropically) favored. Branch 1 is unstable, as also expected from the fact that the specific heat is negative on this branch.

One question of interest is if there is any possible configuration with $T<T_{\rm min}$ and with the same boundary condition as the black hole.
In the case at hand, since we are considering Poincar\' e patch $AdS$, the thermal $AdS$ phase is not available, 
because putting the Poincar\'e $AdS$ geometry at finite temperature introduces a conical singularity.
While in global $AdS$ the  thermodynamics of the system does involve transitions  between 
the black hole geometry and thermal anti-de Sitter space with the same boundary data \cite{Gubser:1998jb, Cvetic:1999ne, Cvetic:1999rb},
this is not possible in the present case.\footnote{Here we correct an earlier version of this paper 
where we  assumed
that at sufficiently low temperatures there would be a phase transition to thermal AdS.
We thank the referee for kindly reminding us of this important point. }
It is possible that the system cannot be cooled below this temperature because there may be no gravitational configuration with lower temperatures:
in trying to cool the system by extracting energy, the system would simply move to the unstable branch with negative specific heat and
increases the temperature again.
Another possibility is that the system might get to lower temperatures by means of more complicated black hole configurations.
In particular, the present model involves other fields, like a charged two-form, thus one would like to see if
there could be black hole configurations with two-form hair that can get to lower temperatures.
In section 4, we will see that this indeed occurs in a modified version of the model where two-form has a general $U(1)$ charge $q$ for
$q$ greater than some critical  value. In that case, we will find that there is a hairy black hole that can get to lower temperatures,
up to a new minimum temperature (lower than $T_{\rm min}$) that depends on the charge $q$. For large $q$, this new minimum temperature goes to 0.

{} The  large temperature  behavior is
\be
\hat s \approx{\pi^2\over 2}\, T^3 + O(T^{-4})\ .
\ee
Hence
\be
C_{\hat\rho } \approx {3\pi^2\over 2}\, T^3+ O(T^{-4})\ .
\label{gars}
\ee
The specific heat diverges as $T\to T_{\rm min}$.

Consider now the behavior of entropy and specific heat near some given low temperature. In particular, near $T_0={3\hat \rho^{1/3}\over 2\pi^{1/3} } $
where the free energy changes sign, we find
\bea
\hat s &\approx & 2\pi\, \hat \rho + 4\pi^{4\over 3}\, \hat\rho^{2\over 3} \ (T-T_0) +\ldots
\nn\\
C_{\hat\rho } &\approx & 4\pi^{4\over 3}\, \hat\rho^{2\over 3} \ T+\ldots\ .
\label{glin}
\eea
The temperature behavior of the specific heat reproduces a standard metal behavior, 
\be
C_{V} \approx c_{\rm e} T+c_{\rm ph}\, T^3
\ee
where the linear and cubic term represent electron and phonon contributions.
At large temperatures, the phonon contribution is dominant and, in the present case, can thus be identified with
$c_{\rm ph}={3\pi^2\over 2}$  (cf. (\ref{gars})). From the linear behavior (\ref{glin}) near the temperature $T_0$, one identifies 
$c_{\rm e}= 4\pi^{4\over 3}\, \hat\rho^{2\over 3}$.


\section{Search for a condensed phase}

Given the existence of a non-zero minimum temperature, an interesting question is what are  the possible gravitational solutions that can
contribute to the thermodynamics below this temperature. We look for charged solutions giving rise to a  finite charge density configuration in the field theory.
One possibility is that there is 
 no stable/metastable ground state for $T<T_{\rm min}$. The Romans theory, however, has more fields and 
 it is possible that the thermodynamically favored solution is actually a black hole configuration 
 with some hair, which will be provided by the extra fields of the supergravity Lagrangian.
Indeed the Lagrangian (\ref{Roman}) contains the complex 2-form $A_{\mu\nu}$ charged under the $U(1)$ symmetry associated with $B_{\mu}$, which 
can in principle  lead to the analogous instability found for the Reissner-Nordstr\" om black hole in \cite{Hartnoll:2008kx,Gubser:2008pf}.
In what follows we shall turn on the 2-form field $A_{\mu\nu}$ and look for such instabilities.
Solutions with $A_{\mu\nu}$ hair can  spontaneously break the $U(1)$ global symmetry of the boundary field theory and 
take the system to a superconducting phase.

To have a more complete understanding on the 2-form dynamics, we consider a slight generalized model with respect to the Romans Lagrangian
in which the 2-form has a general $U(1)$ charge $q$. 
The Lagrangian is given by
\bea
\mathcal{L} &=&\sqrt{g}\ \Big[\ 
R-3X^{-2}\pd_{\m}X\pd^{\mu}X-\frac{X^4}{4}F_{\m\n}F^{\m\n}+\frac{4}{L^2}(X^2+2X^{-1})\Big]
\nn\\
&+&
\frac{L}{8i}\eps^{\m\n\r\s\d}\bar A_{\m\n}\pd_{\r}A_{\s\d}-\frac{q}{8}\eps^{\m\n\r\s\d}\bar A_{\m\n} A_{\r\s}
B_{\d}-\frac{\sqrt{g}}{4X^2}\bar A_{\m\n} A^{\m\n}\ .
\label{GRoman}
\eea
For the value $q=1$ we recover the model (\ref{SRoman}). It is important to notice that 
the STU black hole solutions are also solutions of our generalized model (\ref{GRoman}), since $A_{\m\n}=0$ in the black hole background
and therefore $q$ does not participate in the equations.

The generalization to arbitrary charge $q$ has also the advantage of  permitting the study of the system in the probe limit. This is obtained by
rescaling  $B_{\mu}\rightarrow  B_{\mu}/q$, $A_{\mu\nu}\rightarrow A_{\mu\nu}/q$ and taking the limit $q\to \infty$.
In this limit the Lagrangian for $B_\mu$ and   $A_{\mu\nu}$  decouple from the gravity/dilaton part of the Lagrangian and one can therefore study
the $B_\mu$ and   $A_{\mu\nu}$ system in a fixed $Q=0$ (Schwarzschild anti-de Sitter) background.
This system is of course considerably simpler than the full system that includes the dynamics coupled to $g_{\mu\nu},\ X$.
Nonetheless, since we are also interested in the system at small $q$, in particular, at $q=1$, we will first study the full dynamics
including back reaction, deferring the analysis of the probe limit to section 4.5.


\subsection{The hairy black hole ansatz}

We are interested in finding new black hole solutions with non-trivial profile for the two-form in our generalized gravity theory given by (\ref{GRoman}). The natural ansatz to consider for the metric is

\be\label{setup}
	ds^2=g_{\m\n}dx^{\m}dx^{\n}=e^{2A(r)}\Big(-h(r) dt^2+ dx^2+dy^2+ b(r) dz^2\Big)+e^{2B(r)}\frac{dr}{h(r)}\ , 
\ee	
\be
X=X(r)\ ,\qquad B_{(1)}=\Phi(r)\ dt\ ,\qquad A_{(2)} = A_{(2)} (r)\ .
\ee
As shown below, turning on a non-trivial $A_{(2)} (r)$ necessarily breaks the isotropy  in $x,y,z$. This
is the reason for the introduction of the function $b(r)$, which will be consistent with the choice of direction for $A_{(2)} (r)$  adopted below.\\
Consider the first-order equation for the complex 2-form $A_{(2)}\equiv A_{(2)}^1+iA_{(2)}^2$. Written in components,
\be
	 2i A^{\m\n}= L\, \frac{X^2}{\sqrt{g}}\ \eps^{\m\n\r\s\d}(\pd_{\r}A_{\s\d}-q{i\over L} B_{\r}A_{\s\d})\ .
\label{dosform}
\ee	
The requirement that $A_{(2)}$ is a function only of the radial coordinate leads to the following constraints:
\bea
	&& A^{0r}=0\ ,\\ \nn \\
	&& A^{0i}=\frac{i}{2}\, L\, \frac{X^2}{\sqrt{g}}\ \eps^{0rijk}\ \pd_{r}A_{jk}\ ,\\ \nn\\
	&& A^{ri}=-\frac{q}{2} \, \frac{X^2}{\sqrt{g}}\ \eps^{0rijk}\ B_0A_{jk}\ .\\ \nn
\eea
Due to the antisymmetry of $A_{(2)}$ we can trade $A_{ij}$ by a 3-vector $a_k$ in the three spatial directions,
\be
A_{ij}=\eps_{ijk}a_k\ ,\qquad i,j,k=1,2,3\ .
\label{aqq}
\ee
Next, consider  the Einstein equations (\ref{EqR}). 
The ansatz (\ref{setup}) implies that the Ricci tensor is diagonal, therefore we must choose $A_{ij}$ in such a way that also the stress energy tensor  be diagonal.
A simple choice which is in agreement with (\ref{setup}) is,
\be
\vec{a}=(0,0,a)\ .
\ee
Thus $A_{xy}=a$.  The remaining non-zero components of the $2$-form, $A^{0z}$ and $A^{rz}$, take the form
\be
A^{0z}=i L\, \frac{X^2}{\sqrt{g}}\ \pd_r a\ ,\qquad A^{rz}= - q \frac{X^2}{\sqrt{g}}\ \Phi\ a\ .
\label{app}
\ee
Finally, the equation of motion for $A^{ij}$,
\be
iA^{ij}=- L\, \frac{X^2}{\sqrt{g}}\ \eps^{0rijk}(\pd_{r}A_{0k}+q\frac{i}{L}\, B_0A_{rk})\ ,
\ee
implies the following second-order equation for $a$
\be\label{eqa1}
a''\ +\ \Big(\frac{h'}{h}+\frac{b'}{2b}-B'+\frac{2X'}{X}\Big)a'\ +\ \frac{q^2}{L^2}e^{2B-2A}\frac{\Phi^2}{h^2}a\ -\ \frac{1}{L^2}\frac{e^{2B}}{X^4 h}\ a =0\ .
\ee
Having obtained the equation for the order parameter $a$, we next  obtain  the remaining equations of motion and check the consistency of the ansatz.
In the setup (\ref{setup}) we have explicitly set the fields $B_r$ and $B_i$ to zero; therefore the source terms appearing in their equations of motion must vanish. In particular, from the Maxwell's equations 
\be
\pd_{\m}\Big(\sqrt{g}\ X^4\ F^{\m\r}\Big)=\frac{q}{8}\eps^{\r\a\b\s\d}\bar{A}_{\a\b}A_{\s\d}\ ,
\label{maxi}
\ee
one finds the conditions 
\be
\eps^{k\a\b\s\d}\bar{A}_{\a\b}A_{\s\d}=0\ ,\qquad
\eps^{r\a\b\s\d}\bar{A}_{\a\b}A_{\s\d}=0\ .
\ee
The first one is trivially satisfied because of $A^{0r}=0$; the second condition gives
\be
\eps^{r\a\b\s\d}\bar{A}_{\a\b}A_{\s\d}\ \varpropto\  i(\bar{a}^i\pd_r a_i-a^i\pd_r\bar{a_i})=0\ .
\ee
This is easily satisfied by taking $a\in\mathbb{R}$ (more generally, this is solved by $a=C \tilde a$ with real $\tilde a$ and $C$ is any complex number). 
To complete the analysis of the Maxwell's equations, we write down the equation of motion for the time component of the one-form $B_{(1)}$,
\be\label{eqPhi1}
\Phi''\ +\ \Big(2A'-B'+\frac{b'}{2b}+\frac{4X'}{X}\Big)\Phi'\ -q^2\frac{e^{2B-4A}}{ X^2}\frac{\Phi}{h}a^2=0 \ .
\ee
Now consider  the equation for the scalar field $X$,
\be
\frac{1}{\sqrt{g}}\pd_{\m}\Big(\sqrt{g}\ X^{-1}\pd^{\mu}X\Big)-\frac{X^4}{6}F_{\m\n}F^{\m\n}+X^{-2}\frac{1}{12}A_{\m\n}A^{\m\n}+\frac{4}{3L^2}(X^2-X^{-1})=0\ .
\ee
{}We obtain
\bea
\label{eqX1}
&& X''+\Big(4A'-B'+\frac{h'}{h}+\frac{b'}{2b}-\frac{X'}{X}\Big)X'+\frac{X^5e^{-2A}}{3h}\Phi'^2
\nn\\
&& +\frac{e^{2B-4A}}{6Xh}a^2-\frac{L^2}{6}X^3e^{-4A}\Big(a'^2-\frac{q^2}{L^2}e^{2B-2A}\frac{\Phi^2a^2}{h^2}\Big)+\frac{e^{2B}}{h}\frac{4}{3L^2}(X^3-1)=0\ .
\eea
Finally,  the Einstein equations read
\begin{equation}
R_{\mu\nu}-{1\over 2}\, g_{\mu\nu}R\, =T_{\mu\nu}\ ,
\end{equation}
with
\begin{eqnarray}
\label{stressenergy}
T_{\mu\nu}
&=& \sqrt{g}\ \Big[\ 
-3X^{-2}\pd_{\m}X\pd_{\nu}X-\frac{X^4}{2}F_{\m\rho}F_\n^{\ \ \rho} -\frac{1}{2\,X^2}\, \bar A_{\m\rho}\, A_{\n}^{\ \ \rho}
\Big]\\ \nonumber&-&\frac{1}{2}\,\sqrt{g}\, g_{\m\n}\, \Big[\ 
-3X^{-2}\pd_{\m}X\pd^{\mu}X-\frac{X^4}{4}F_{\m\n}F^{\m\n}-\frac{1}{4\, X^2}\, \bar A_{\m\n}\, A^{\m\n}+\frac{4}{L^2}(X^2+2X^{-1})\Big].
\end{eqnarray}
In order to fix the metric equations of motion, we consider the three linear combinations
\be
R^t_t-R^x_x\ ,\qquad R^z_z-R^x_x\ ,\qquad R^r_r-R^t_t-R^z_z-2R^x_x\ , 
\ee
and the $xx$ equation. We find
\be\label{eqh1}
h''+\Big(4A'-B'+\frac{b'}{2b}\Big)h'=X^4e^{-2A}\Phi'^2+L^2e^{-4A}X^2ha'^2+\frac{e^{2B-4A}}{X^2}a^2\ ,
\ee
\be\nn
A''+\Big(4A'-B'+\frac{h'}{h}+\frac{b'}{2b}\Big)A'=
\ee
\be\label{eqA1}
=\frac{4}{3L^2}e^{2B}\frac{(X^2+2X^{-1})}{h}-\frac{X^4e^{-2A}}{6h}\Phi'^2-\frac{e^{2B-4A}}{3X^2h}a^2-
\frac{L^2}{6}X^2e^{-4A}\Big(a'^2-\frac{q^2}{L^2}e^{2B-2A}\frac{\Phi^2a^2}{h^2}\Big)\ ,
\ee
\\
\be\label{eqb1}
\frac{b''}{b}+\Big(4A'-B'+\frac{h'}{h}-\frac{b'}{2b}\Big)\frac{b'}{b}=L^2X^2e^{-4A}\Big(a'^2-\frac{q^2}{L^2}e^{2B-2A}\frac{\Phi^2a^2}{h^2}\Big)+
\frac{e^{2B-4A}}{X^2}\frac{a^2}{h}\ ,
\ee
\\
and a first-order constraint
\\
\be\label{eqConstr}
12A'^2+3A'\frac{h'}{h}+\Big(3A'+\frac{h'}{2h}\Big)\frac{b'}{b}+\frac{1}{2}\frac{e^{2B-4A}}{X^2}\frac{a^2}{h}+\nn
\ee
\be
+X^4e^{-2A}\frac{\Phi'^2}{2h}-\frac{L^2}{2}X^2e^{-4A}\Big(a'^2-\frac{q^2}{L^2}e^{2B-2A}\frac{\Phi^2a^2}{h^2}\Big)-3\frac{X'^2}{X^2}-
\frac{4}{L^2}e^{2B}\frac{X^2+2X^{-1}}{h}=0\ .
\ee
\\
The freedom of radial coordinate redefinitions will be fixed by the choice  $e^{B}=e^{-2A}r$, i.e. the same condition obeyed by the ``bald" charged black hole (\ref{atun}). 
The constraint (\ref{eqConstr}) can be used to simplify the r.h.s. of the second order equation for $A$:
\\
\be\label{EEqA}
A''-\Big(B'+\frac{b'}{2b}\Big)A'+\frac{X'^2}{X^2}-\frac{1}{6}\frac{h'}{h}\frac{b'}{b}+\frac{L^2}{3}e^{-4A}X^2a'^2+\frac{e^{2B-4A}a^2}{6X^2h}=0\ .
\ee

\subsection{Symmetries and conserved charges}

The equations of motions (\ref{eqa1}), (\ref{eqPhi1}), (\ref{eqX1}) and (\ref{eqh1})-(\ref{eqConstr}) can be obtained from the Effective Lagrangian,
\bea\label{Leff}
\mathcal{L}_{\rm eff}&=&h\sqrt{b}e^{4A-B}\Big[12A'^2+3A'\frac{h'}{h}+3A'\frac{b'}{b}+\frac{1}{2}\frac{h'}{h}\frac{b'}{b}-3\frac{X'^2}{X^2}\Big]
\nn\\
& &+\ \frac{1}{2}X^4\sqrt{b}e^{2A-B}\Phi'^2+\ \frac{4}{L^2}\sqrt{b}e^{4A+B}(X^2+2X^{-1}) 
\nn\\
& &-\ \frac{L^2}{2}X^2e^{-B}h\sqrt{b}a'^2+\ q^2\frac{X^2}{2}e^{-2A+B}\frac{\sqrt{b}}{h}\Phi^2a^2-\ \frac{\sqrt{b}}{2X^2}e^Ba^2.
\eea
We will now search for symmetries under scaling transformations,
\be
t\rightarrow\lambda_t t\ ,\qquad (x,y)\rightarrow\lambda_{(x,y)}\ (x,y)\ ,\qquad z\rightarrow\lambda_z z\ , 
\ee
with $\lambda=1+\delta \lambda $, together with the associated infinitesimal transformations for the fields,
\bea
&& h\rightarrow(1+\eps_h)h    \ ,\qquad  A\rightarrow A+\eps_A \ , \qquad B\rightarrow B+\eps_B\ ,
\nn\\ 
&& \Phi\rightarrow(1+\eps_{\Phi})\Phi \ ,\ \ \  b\rightarrow(1+\eps_b)b\ ,\ \ \
a\rightarrow(1+\eps_{a})a\ .
\eea
Because  $V(X)$ is not an homogeneous polynomial the dilaton $X$ must scale trivially, $X\to X$. We now demand that 
${L}_{\rm eff}$ and the background  are invariant 
under the above scaling transformations:
\be
\delta\mathcal{L}_{\rm eff} =\delta{ds^2}=\delta{A_{(2)}}=\delta{B_{(1)}}=0\ .
\ee
This leads to an algebraic system admitting a two-parameter family of solutions. Choosing $\eps_A$ and $\eps_b$ as independent 
parameters, we find
\be\label{scaling1}
\phantom{AAA}
\begin{array}{lll}
\eps_h= -8\eps_A - \eps_b\ , \qquad \eps_B=-4\eps_A - {\eps_b\over 2}\ ,\\  \\
\eps_{\Phi}= 3\eps_A -{\eps_b\over 2}\ ,\qquad \eps_a=2\eps_A\ ,\\ \\
\delta\lambda_t =3\eps_A +{\eps_b\over 2}\ ,\qquad \delta\lambda_z=-\eps_A -{\eps_b\over 2}\ ,\qquad \delta\lambda_{(x,y)}= -\eps_A\ .
\end{array}
\ee 
The transformations with parameters (\ref{scaling1}) represent two scaling symmetries of the Lagrangian (\ref{Leff}) (before eliminating $B$  by a choice of radial coordinate). 
The two scaling symmetries are summarized in table \ref{scaling2}, where $\vec x=(x,y,z)$ and the charges $\alpha$ are assigned following the rule $\mathfrak{f}\rightarrow\lambda^{\alpha}\mathfrak{f}\ $, for a generic field or variable $\mathfrak{f}$.

\begin{table}[htbp] \label{scaling2}
\centering
\begin{tabular}{c|c|c|c|c|c|c|c|c}
	{\rm Symm.}&\ t\ &\ $\vec x$ & $r$ & $e^{A}$ &\ $h$ & $X$ &\ $\Phi$ & $ a$ \\
	\hline\hline
	\rule{0pt}{4mm}{\rm I}&-1&-1&1&1&\phantom{-}0&0&\phantom{-}1&2\\
	{\rm II}              &\phantom{-}1&-1&0&1&-4&0&-1&2\\
	\end{tabular}
\caption{ Weights for the scaling symmetries of the effective Lagrangian (\ref{Leff}).}
\end{table}

Using the Noether procedure we find the two associated conserved charges,
\bea
&&Q_1=h\sqrt{b}e^{4A-B}\left(3\frac{h'}{h}-\frac{b'}{b}\right)-2X^2L^2h\sqrt{b} e^{-B}\ a'a-  3\sqrt{b}X^4 \ e^{2A-B}\  \Phi'\Phi,\qquad
\nonumber\\
&&Q_2=h\sqrt{b}e^{4A-B}\left(\frac{h'}{h}-\frac{b'}{b}\right)-\sqrt{b}X^4 \ e^{2A-B}\ \Phi'\Phi\ .
\eea
It is easy to check that by differentiating these equations one obtains a linear combination of the differential equations
of the system given in section 3.1; in other words, the charges $Q_1, \ Q_2$ represent two integrals of the equations of motion.
Note the combination
\be
Q_3\equiv Q_1-3Q_2 = 2 h\sqrt{b}e^{4A-B} \ \frac{b'}{b} -2X^2L^2h\sqrt{b} e^{-B}\ a'a\ ,
\ee
which exhibits the fact that a non-trivial $a(r)$ turns on the metric component $b(r)$.

\section{Numerical analysis}

The numerical problem involves the resolution of six coupled second-order differential equations from the black hole horizon up to infinity, 
where boundary conditions need to be fixed by the ``shooting" method, which means that one needs to satisfy boundary requirements both at infinity and at the horizon.
The strategy is a slight generalization of the  procedure explained in detail in \cite{Hartnoll:2008kx,Gubser:2008pf}, which we here review, 
with emphasis on the new aspects we encountered that are inherent to the present system.\newline

\subsection{Boundary conditions} 

We look for black hole solutions 
with regular event horizons. The location $r=r_h$ of the horizon is defined by the
simple zero of $h$ lying at larger $r$.
The Hawking temperature associated with  black holes of the form (\ref{setup}) is then given by
\be\label{Temp}
T_{Hawk}=\frac{1}{4\pi}\frac{e^{3A(r_h)}h'(r_h)}{r_h}\ .
\ee
Regularity of the horizon requires that $h'(r_h)$ is a non-zero finite quantity.

As a warm-up example, let us first consider the  numerical derivation of the bald black hole (\ref{atun}), which has $a=0$. 
In this case we can consistently set the metric component $b=1$.
We fix the horizon coordinate at $r_h=1$ by means of symmetry I. Then the remaining four equations 
(\ref{eqPhi1}), (\ref{eqX1}), (\ref{eqh1}), (\ref{EEqA}), being second order, are completely specified by eight boundary values at $r_h$. 
Not all of them are free parameters: in order to have a fully regular solution, one must require that $\Phi(r_h)=0$ and that $X$ is regular at the horizon. This condition gives
\be\label{Xhorizon}
X'(r_h)=\left[\ \frac{4}{3}\frac{e^{2B}(1-X^3)}{h'}-\frac{X^5e^{-2A}}{3 h'}\Phi'^2\right]_{r=r_h}\ .
\ee  
Furthermore, one integration constant is eliminated by the energy constraint and another 
integration constant can be fixed by symmetry II. 
Therefore, based on the horizon boundary conditions, there exists a three-parameter family of solutions. 
This means that two additional constraints have to be imposed in order to match the analytic expression of the bald black hole, which, at fixed charge density,
depends on a single parameter --which can be taken to be the temperature.
Let us now consider the asymptotic behavior. {}From the fact that the metric approaches the anti-de Sitter solution, we obtain 
\bea
&&A=A_{\infty}+\log\ {r\over L}\ + \frac{Q_A}{r^2}+\ldots\label{InftyNBaldA}\\ \nn \\
&&h=h_{\infty}-\frac{\mm L^2}{r^4}+\ldots\label{InftyNBaldh}\\ \nn \\
&&\Phi=\mu L-\frac{4\pi^2\hat \rho L^5}{r^2}+\ldots\label{InftyNBaldPhi}\\ \nn \\
&&X=1+\frac{Q_X}{r^2}+\frac{C_X}{r^2}\log\ {r\over L}+\ldots\label{InftyNBaldX}
\eea
 We now require that $C_X=0$ and $Q_X=2Q_A\ $. 
The first requirement removes logarithmic terms; the  second requirement fixes the dilaton charge to a special value 
(which allows one to find an analytic solution, see \cite{Behrndt:1998jd}).
These conditions translate into two non-linear relations on the three horizon parameters, leaving only one free parameter.
Then numerical integration of the differential equations reproduces the bald black hole  solution (\ref{atun}).
\newline 

It is now
straightforward to apply the same procedure for black holes with hair. 
{}For the hairy black hole ansatz, 
the condition (\ref{Xhorizon}) gets modified by the addition of the term $-{e^{2B-4A}a^2}/(6Xh')$ evaluated at $r_h$. 
Now consider  the second-order equations of motion (\ref{eqa1}), (\ref{eqb1}) for $a$ and $b$. We are adding four more boundary values at $r_h$ to the previous discussion.\footnote{Here one may also
use
the conserved charges discussed above to write a first order equation for $b$. This leads to equivalent results.}
The regularity conditions
\be
a'(r_h)=\left[\frac{e^{2B}a}{X^4h'}\right]_{r=r_h}\ ,\qquad b'(r_h)=\left[\frac{e^{2B-4A}a^2}{X^2h'}\right]_{r=r_h}\ ,
\ee
and the rescaling $b\rightarrow\lambda b$, reduce these four parameters  to a single one. 
An additional requirement comes from the physics we aim to describe, namely a background that represents
spontaneous $U(1)$ symmetry breaking in the dual field theory.
This proceeds as usual: the 2-form $A_{(2)}$ is dual to the operator $O$ that is expected to condense
(for a discussion on the dual operator, see section 5).
Therefore, from the asymptotic behavior of $a$,
\be
\label{ainfty}
a=O_1r+\frac{O_2}{r} +\ldots\ ,
\ee
the coefficient of the non-normalizable term, $O_1$, is interpreted, in the field theory,  as a source term for the operator $O$, 
whereas the coefficient of the normalizable term, $O_2$, gives the value of the condensate.
Then, demanding {\it spontaneous} (rather than explicit) symmetry breaking of the global $U(1)$ in the field theory amounts to imposing $O_1=0$.
This condition fixes the additional parameter we found in considering the equations (\ref{eqa1}), (\ref{eqb1}). 

In the case that $O_1$ is different from zero, the value of the condensate is affected by logarithmic divergences, originating from a term
$\frac{n_1}{r}\log r$ in the asymptotic behavior that is not shown explicitly in (\ref{ainfty}).
These logarithmic terms disappear upon our choice $O_1=0$.

\medskip

Summarizing, we obtain  a one-parameter family 
of superconducting black holes, provided there {\it are} solutions satisfying this condition $O_1$, which, as we will see, it is not always the case.
In particular, For the Romans Lagrangian (\ref{Roman}), we have studied the coupled system of six differential equations numerically and found no solution which
obeys the required boundary condition. 
Indeed, a detailed study shows that any solution which is regular at the horizon has non-vanishing $O_1$ and therefore does not represent $U(1)$ spontaneous symmetry breaking. The underlying reason will be understood in the next subsection: the charge $q=1$ of the $A_{\mu\nu}$ field is not sufficiently large to drive to an instability.

%

\medskip

Some useful information can be obtained by computing the conserved charges both at the horizon and at infinity.
Because $h$ vanishes at the horizon, we find that $Q_3=0$.
Now computing  $Q_3$ at infinity and using that  $O_1=0$, we find that $b=1+O(1/r^6)$.
Now consider  the calculation of $Q_1$. At the horizon we find
$Q_1=4\pi^2T_H \hat s$, where $\hat s$ is the entropy density normalized as in section 2.
Now computing $Q_1$  at infinity we find $Q_1=4\pi^2 ({4\over 3}\hat \epsilon-\mu\hat\rho )$.
Thus charge conservation implies 
the thermodynamic relation
\be
\hat\epsilon  = {3\over 4} (T\hat s+\mu\hat\rho )\ .
\label{ther}
\ee
In \cite{Gubser:2009cg}, in a different context, it was noticed that these type of relations follow from the assumption of a traceless field-theory energy-momentum tensor.
In the present, four-dimensional case, this assumption implies $\hat\epsilon=3\hat p$, where $\hat p $ is the pressure. Now one uses the relation
$-\hat p=\hat \epsilon-T_H\hat s-\mu\hat \rho$, thereby, strikingly, the relation (\ref{ther}) follows.


\subsection{Critical temperatures}

Consider the theory given by the Lagrangian (\ref{GRoman}), where the charge $q$ is taken as a real parameter. Before presenting the full analysis including back-reaction, we first obtain the curve of critical temperature as function of the $q$ parameter.

The idea is based on the following observation. In second (or higher) order phase transitions, the order parameter approaches zero near the critical temperature.
In the gravity solution, $O_2\rightarrow 0$ implies that $a\rightarrow 0$, which in turn implies
that the hairy black hole  approaches the bald black hole (\ref{atun}). 
Therefore, in the vicinity of a continuous phase transition, we just need to study the $a$ equation (\ref{eqa1}) in the bald black hole background (\ref{atun}). This has
\be
X=H^{1/3}\ ,\qquad b=1\ .
\ee
While this method is numerically very accurate for the determination of the critical temperature, it is however 
inappropriate to detect possible first-order phase transitions, where $a$ is {\it not} small near
the transition. The complete picture that covers the case of first-order transitions as well
will be clear upon solving the full system including back reaction.


Substituting the bald black hole solution in (\ref{eqa1}) we find
\be\label{eqetan}
a''\ +\ {r^4-Q^2 r^2+3\bar m\over r^3\td f}   a'\ +{1\over \td f\, (r^2+Q^2)}\Big( {\bar m q^2 Q^2 (r^2-r_h^2)^2\over r^2\td f\, (r_h^2+Q^2)^2} -r^2   \Big) a\ \ =0\ ,
\ee
with
\be
\td f \equiv r^2+Q^2-\frac{\bar m}{r^2}\ ,\qquad \mm \equiv \bar m/L^2 \ . 
\ee
By a further rescaling $\bar m = r_h^4 \tilde m$, $Q=r_h \tilde Q$ and introducing a new variable $z=r_h/r$, the dependence on $r_h$ drops out from the equation. In addition, we note that
 $\tilde m =1+\tilde Q^2$ by virtue of the horizon equation. The final equation depends only on the parameters $\tilde Q$ and $q$.
Define $p(z) =  z\ a(z)$, so that, at small $z$ (large $r$), $p$ has the expansion
$$
p(z) = r_h\ O_1+ O_2 \frac{z^2}{r_h}+n_1\frac{z^2}{r_h}\log z +....
$$ 
The differential equation becomes
\be
p''(z)+F(z) p'(z)+G(z)p(z)=0\ ,
\label{tyu}
\ee
\bea
&& F(z)\equiv \frac{3 \left(\tilde Q^2+1\right) z^5-\tilde Q^2
   z^3+z}{z^2 \left(z^2-1\right) \left(\left(\tilde Q^2+1\right)
   z^2+1\right)}  
\nonumber\\
&& 
G(z)\equiv \frac{ q^2 \tilde Q^2 z^2
   \left(z^2-1\right)-\left(\tilde Q^2+1\right) z^4
   \left(\left(\tilde Q^2+1\right) z^2+1\right) \left(\left(3
   z^2-1\right) \tilde Q^4+3 \left(z^2+1\right)
   \tilde Q^2+3\right)}{\left(\tilde Q^2+1\right) z^2
   \left(z^2-1\right) \left(\tilde Q^2 z^2+1\right)
   \left(\left(\tilde Q^2+1\right) z^2+1\right)^2}
\nonumber
\eea
Regularity near the horizon (located at $z=1$) implies the boundary condition
\be
p'(1)=\frac{\left(2 \tilde Q^4+6 \tilde Q^2+3\right) p(1) }{2
   \left(\tilde Q^4+3 \tilde Q^2+2\right)}\ .
   \ee

As explained in section 4.1, we must look for solutions which have $O_1=p(z=0)=0$.
Solutions with $O_1=0$ have no logarithmic terms in the expansion at infinity, in particular, $n_1=0$.
In general, one has $O_1=O_1(\tilde Q,q)$ hence $O_1=0$ gives $\tilde Q=\tilde Q(q)$. 
The critical temperature is then obtained from the Hawking temperature (\ref{hawk}) which, expressed 
in terms of the new variables, reads
\be
T=\frac{ \left(\tilde Q^2+2\right) r_h}{2 \pi  L^2
   \sqrt{\tilde Q^2+1}}=  \frac{ \left(\tilde Q^2+2\right) \hat\rho^{1/3}}{2 \pi  L
   \tilde Q^{1/3} (\tilde Q^2+1)^{2/3}} \ .
\ee
Thus we have $T=T(q)=T\big( \tilde Q(q)\big)$.
Figure 2 shows the temperature as a function of the $A_{\mu\nu}$-charge $q$ obtained from the numerical resolution of the differential equation
(\ref{tyu}) (we set $\hat \rho=1$).
We see that solutions with $O_1=0$ exist for $q$ above some critical value, $q_{\rm cr}\approx 5.407$. 
For $q>q_{\rm cr}\approx 5.407$ there are two values of $\tilde Q$ that solve $p_1(\tilde Q,q)=0$ 
and hence two values for the temperature $T_1(q)\le T_2(q)$.
We anticipate the result that the set of physically relevant critical temperatures will be the higher branch $T=T_2(q)$.
\footnote{For higher $q$, new branches appear, but they are not physically relevant as the corresponding critical temperatures are, again, smaller than $T_2(q)$.} 
More features of this curve will be discussed in the next subsection.

\smallskip

\begin{figure}[h!]
\centering
\includegraphics[scale=0.5]{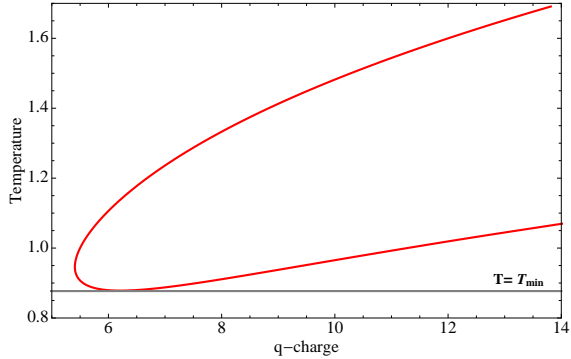} 
\caption{Critical temperatures as a function of the $A_{\mu\nu}$-charge $q$.}
\label{RomansTc}
\end{figure}

\subsection{Numerical analysis of the full system including back-reaction}

The theory (\ref{GRoman}) contains different black hole solutions. The uncondensed $a=0$ phase is described as usual by the bald solution (\ref{atun}). 
When the temperature is below the critical value $T_2(q)$ (upper branch in fig. \ref{RomansTc}), one or more hairy black hole solutions appear, depending on the value of the parameter $q$. 
We have numerically solved the six coupled differential equations obtained in section 3.1 --corresponding to the ansatz (\ref{setup})--
with the boundary conditions as described in section 4.1. The numerical calculation is done at fixed charge density $\hat \rho =1$.

\begin{figure}[htbp] 
\centering%
\subfigure[]%
{\includegraphics[scale=.45]{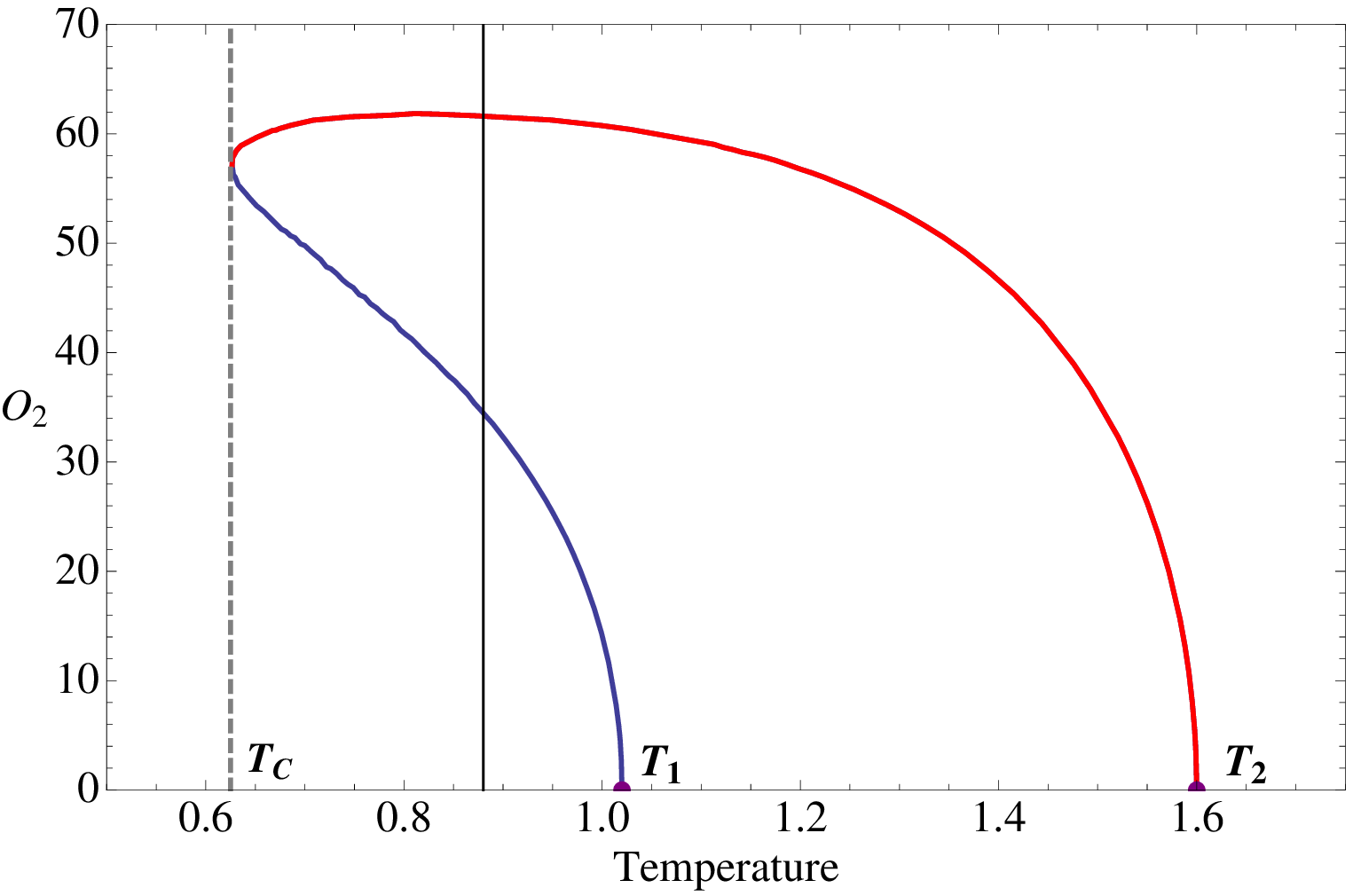}}\hspace{.6cm}
\subfigure[]%
{\includegraphics[scale=.45]{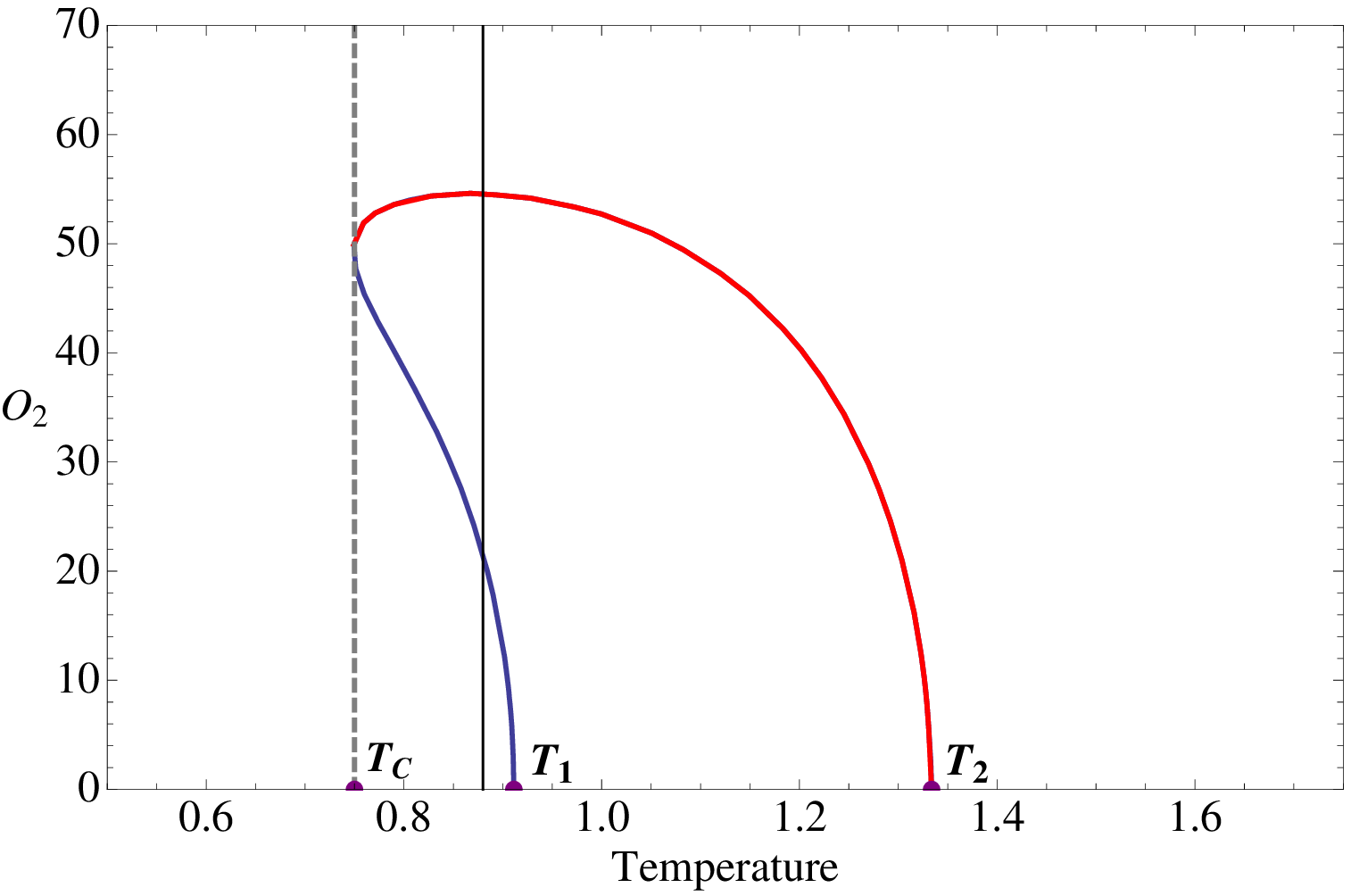}}
\caption{(a) and (b) show  $O_2(T)$ for $q=12$ and $q=8$, respectively. Only the red part (upper branch to the right of vertical dashed line) of the  plots has a physical relevance: in that range of temperatures the 
system will be in the condensed phase.  The solid vertical line corresponds to the $T_{\rm min}$ of the bald black hole. Since $T_C<T_{\rm min}$, 
the system can reach  temperatures lower than $T_{\rm min}$ through the hairy black hole configuration.}
\label{CondSol1}
\end{figure}

\begin{figure}[htbp] 
\centering%
\includegraphics[scale=.5]{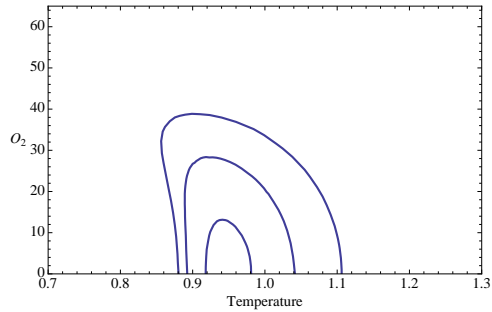}
\caption{Condensate $O_2(T)$ for the values $q=6$, $q=5.65$, $q=5.45$ (from top to bottom).}
\label{CondSol2}
\end{figure}

Figure \ref{CondSol1}a shows the condensate $O_2$ as a function of the temperature for $q=12$. Note that for  $T_C\approx 0.62<T<T_1  $ there are two black hole solutions with $A_{\mu\nu}$ hair. 
Figure \ref{CondSol1}b is the similar plot for $q=8$. 
The existence of a range of temperatures with two black hole solutions is a feature which is common 
for $q\gtrsim 6$, as can be seen from figure \ref{CondSol2}.  Figure \ref{CondSol2} also illustrates the fact 
that the maximum of $O_2(T)$ is going to zero when $q$ is reaching $q_{\rm cr}$. Therefore the hair of the black hole solutions continuously disappear as $q\to q_{\rm cr}$. 
Figures \ref{CondSol1}, \ref{CondSol2} also show that  the condensed phase gets extended to lower temperatures as $q$ is increased.
A natural guess is that in the limit $q\to\infty $ the condensed phase extends all the way down to $T=0$.
This guess will be supported by the study of the probe limit in section 4.5.
For finite $q$, $T_C$ represents the minimum temperature that the system can reach.
This is a similar situation to the one explained in section 2: if one attempts to cool the system by extracting energy, the system will be pushed to the
unstable branch, resulting into an increase of temperature (unless a more complicated gravitational configuration with $T<T_C$  exists that could provide a new
equilibrium configuration).

\subsection{Free energy and Phase Diagram}

In the canonical ensemble the configuration that dominates the thermodynamics is the one with least Helmholtz free energy 
$F$.
We have compared the free energies of two different phases corresponding to the bald black hole and the hairy black hole.
For $T>T_2(q)$ there is no hairy black hole solution and the thermodynamically 
relevant solution is the bald black hole (\ref{atun}).
The hairy black hole solution appears at $T<T_2(q)$. We have numerically studied the free energy of the hairy black holes 
for different values of the 2-form charge $q$. The calculation proceeds as follows. From the formula
$$
F=\hat{\epsilon}-T\hat{s} \ ,
$$
we see that calculating the free energy of the hairy solutions  requires a combination of asymptotic and horizon quantities. 
We first obtain the value of $m$ from the expansion (\ref{InftyNBaldh}); then the energy can be deduced using (\ref{TermoQuant})
and the entropy from the standard definition in terms of the area of the horizon.
The   temperature is computed by using (\ref{Temp}).

We begin by  discussing what happens for a fixed value of $q$. From fig. \ref{CondSol2} we see that for $q>5.65$ there is a region of temperatures
where there are two black holes at the same temperature.
In general, we find that the   lower branch solution has a  free energy which is always higher than the free energy of the upper branch solution.

\begin{figure}[htbp] 
\centering%
{\includegraphics[scale=.41]{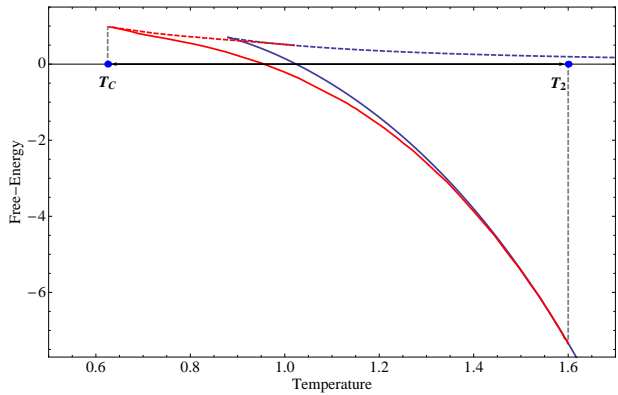}}\hspace{1cm}
\caption{ Free energies for the bald (blue, upper curve) and hairy (red, lower curve) black holes at $q=12$. 
}
\label{FreeRoman}
\end{figure}

Figure \ref{FreeRoman} shows the  free energy  for the bald black hole (upper, blue curve)  and for the hairy black hole 
(lower, red curve) for  $q=12$.
In the range of temperatures $T_C\le T\le T_2$, the hairy black hole solution has less free energy with respect to the uncondensed black hole configuration, therefore  the system stays in the condensed phase. The transition at $T=T_2$ from
the uncondensed to the condensed phase is second-order, since the first derivative of the free energy is continuous.

\begin{figure}[h!]
\centering
\subfigure[]%
{\includegraphics[scale=0.4]{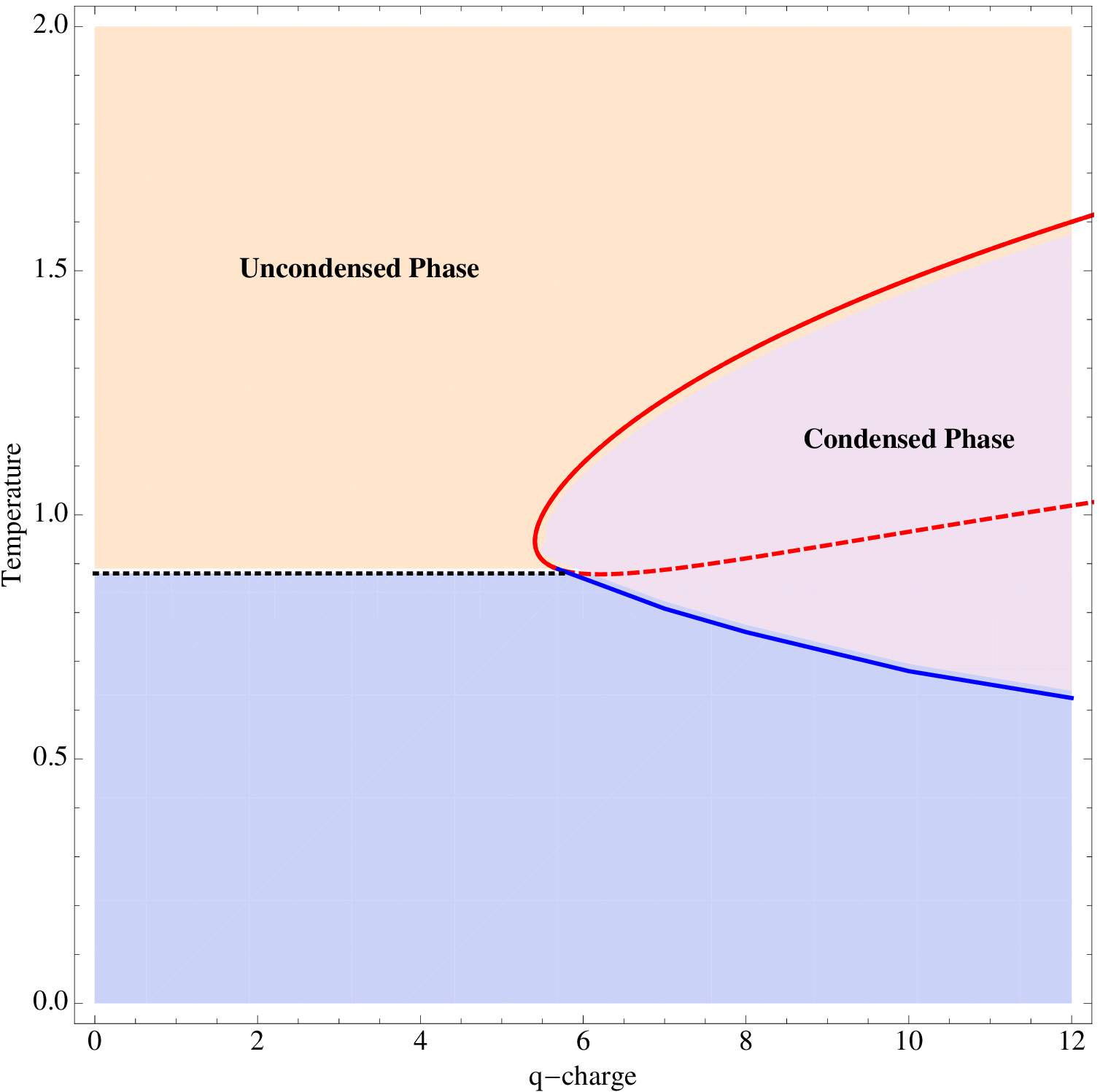}} 
\hspace{.6cm}
\subfigure[]%
{\includegraphics[scale=0.4]{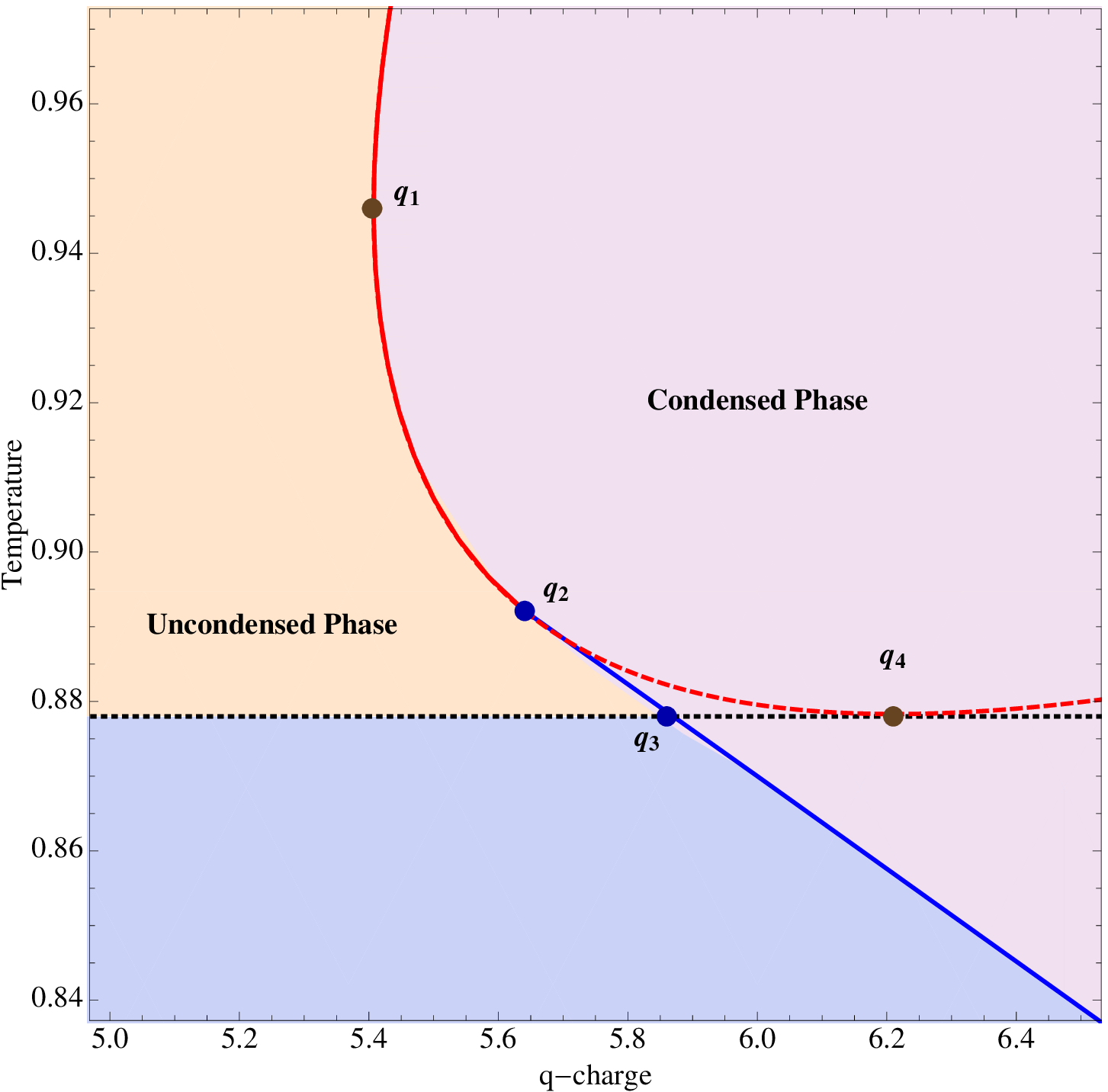} }
\caption{(a) Phase Diagram for the system (\ref{GRoman}). (b) Zoom-in of the same diagram.
The dashed line after the point 2 is the remaining part of the curve in figure 2, that has now become physically irrelevant as does not represent
any separation between phases. The dotted horizontal line represents the minimal temperature of the bald black hole.}
\label{PhDiagr}
\end{figure}

The main result of this section is summarized by the phase diagram shown in Figures \ref{PhDiagr}(a),(b).
Recall that the minimum temperature (\ref{Tmini}) of the bald black hole is $T_{\rm min}\approx 0.878\approx 0.88$
and that $T_1(q)$, $T_2(q)$ denote the lower and upper branches in figure \ref{RomansTc}. 
There are four special points in this diagram, summarized in table 2:

\begin{enumerate}

\item The minimum charge at which the hairy black hole solution exists is  $q=q_1\equiv q_{\rm cr}\approx 5.407$, at which $T\approx  0.946$.

\item The critical charge $q_2 \approx 5.65 $ such that for $q>q_2$  there 
are two hairy black hole solutions with the same $T$ in a range of temperatures.

\item A  $q=q_3\approx 5.85$ at which $T_C$ coincides with $T_{\rm min}$. For $q<q_3$ one has $T_{\rm min} <T_C$; for  $q>q_3$ one has $T_{\rm min} >T_C$.

\item 
The minimum   temperature  on the branch $T_1(q)$. This is $T\approx 0.878$. At this temperature,  $q=q_4\approx 6.211$. 
Note that this exactly coincides with the minimum temperature of the bald black hole solution, $T_{\rm min}$. 

\end{enumerate}

\begin{table}[htbp]\label{qvsT}
\centering
\begin{tabular}{c|c|c|}
& $q$ & $T$ \\
\hline\hline
\rule{0pt}{4mm}
1&5.41 & 0.946 \\
2&5.65 & 0.892\\
3&5.85 & 0.878 \\
4&6.21 & 0.878 \\
\end{tabular}
\caption{Values of  charges  and critical temperatures at the four special points of figure  \ref{PhDiagr}.}
\end{table} 

Summarizing, coming from high temperatures, one finds the following phase transitions:

\smallskip

\noindent - For $q<q_{\rm cr}\equiv q_1$, the system stays all the way down to $T_{\rm min}$ in the uncondensed phase described by the bald black hole.

\smallskip

\noindent - For $q_1<q<q_2$, the system undergoes a second-order phase transition from the uncondensed to the condensed (superconducting) phase at $T_2(q)$ and when
the temperature reaches $T_1(q)$ there is another second-order phase transition back to the uncondensed phase (an example is $q=5.45$ in fig. \ref{CondSol2}).
It stays there until $T_{\rm min}$.

\smallskip

\noindent - For $q_2<q<q_3$,  the system undergoes a second-order phase transition from the uncondensed to the condensed (superconducting) phase at $T_2(q)$, where it stays until
the temperature reaches the minimum temperature $T_C$ of the hairy solution. In the range $T_{\rm min}<T<T_C$, the bald black hole is the only remaining solution and the system
should undergo a transition back to the uncondensed phase.

\smallskip

\noindent - For $q>q_3$,  the system undergoes a second-order phase transition from the uncondensed to the condensed (superconducting) phase at $T_2(q)$ and stays there until it reaches the
minimum temperature $T_C$ (below which there is probably no equilibrium configuration).

In the field theory description,  the dilatonic black hole
represents a metallic (uncondensed) phase. This is supported by the temperature behavior of the entropy and
specific heat as discussed in section 2.2.
The hairy black hole represents a superconducting 
 phase.

\subsection{Probe limit}

\begin{figure}[h!]
\centering
\includegraphics[scale=1]{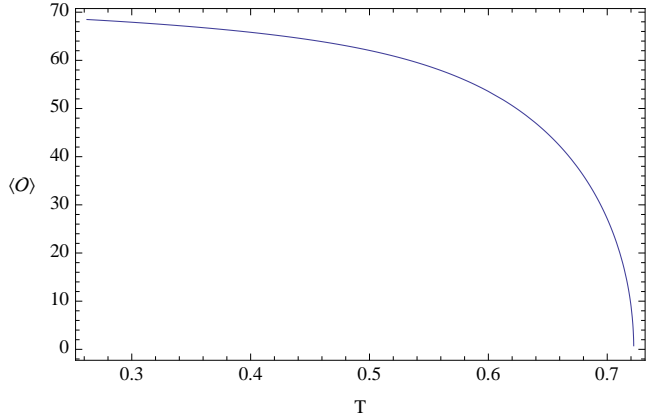} 
\caption{Condensate as a function of the temperature in the probe approximation.}
\label{condTprobe}
\end{figure}

As pointed out in section 3, the introduction of the charge parameter $q$ has the additional bonus of granting a probe limit, 
whereby the relevant dynamics for condensation is encoded in a decoupled non-gravitational sector. To that matter, let us consider our stress-energy tensor for the complete system

\begin{eqnarray}
T_{\mu\nu}
&=& \sqrt{g}\ \Big[\ 
-3X^{-2}\pd_{\m}X\pd_{\nu}X-\frac{X^4}{2}F_{\m\rho}F_\n^{\ \ \rho} -\frac{1}{2\,X^2}\, \bar A_{\m\rho}\, A_{\n}^{\ \ \rho}
\Big]\\ \nonumber&-&\frac{1}{2}\,\sqrt{g}\, g_{\m\n}\, \Big[\ 
-3X^{-2}\pd_{\m}X\pd^{\mu}X-\frac{X^4}{4}F_{\m\n}F^{\m\n}-\frac{1}{4\, X^2}\, \bar A_{\m\n}\, A^{\m\n}+\frac{4}{L^2}(X^2+2X^{-1})\Big].
\end{eqnarray}
Then we consider a limit in which we re-scale

\begin{equation}
B_{\mu}\rightarrow q^{-1}\, B_{\mu}\ ,\qquad A_{\mu\nu}\rightarrow q^{-1}\, A_{\mu\nu}\ .
\end{equation}
The $\{B_\mu,\, A_{\mu\nu}\}$ part of the action scales homogeneously as $q^{-2}$.
In the $q\rightarrow \infty$ limit the $ B_\mu$ and $ A_{\mu\nu}$       do not back-react on the geometry and the gravitational sector 
gives rise to a Schwarzschild-Anti de Sitter geometry (where $X=1$).
 The $\{B_\mu,\, A_{\mu\nu}\}$  sector  can then be studied independently. 

\begin{equation}
\label{Lp}
\mathcal{L} =-\sqrt{g}\, \frac{1}{4}F_{\mu\nu}F^{\mu\nu}-\frac{\sqrt{g}}{4} \bar A_{\mu\nu}A^{\mu\nu}+
\frac{L}{8i}\epsilon^{\mu\nu\rho\sigma\delta}\bar A_{\mu\nu}\partial_{\rho}A_{\sigma\delta}-\frac{1}{8}\epsilon^{\mu\nu\rho\sigma\delta}\bar A_{\mu\nu}A_{\rho\sigma}B_{\delta}\ ,
\end{equation}
in the background of the Schwarzschild-AdS black hole, which reads

\begin{equation}
ds^2=-f\, dt^2+\frac{dr^2}{f}+\frac{r^2}{L^2}\, d\vec{x}^2,\qquad f=\frac{r^2}{L^2}-\frac{M^2}{r^2}\ .
\end{equation}
This geometry has a horizon at $r_h=\sqrt{M\, L}$. The horizon radius is related to the temperature as

\begin{equation}
T=\frac{r_h}{\pi\,L^2}\ .
\end{equation}
We can read off the equations of motion from the generic ones in (\ref{eqa1}) and (\ref{eqPhi1}). Introducing a new variable $z=\frac{r_h}{r}$ and  redefining

\begin{equation}
B_0=\Phi=\frac{r_h}{L}\,\varphi\qquad a=\frac{r_h^2}{L^3}\,\frac{p}{z}\ .
\end{equation}
the differential equations to solve become

\bea
&&z\,\varphi''-\varphi'-\frac{z}{1-z^4}\, p^2\,\varphi=0 \ ,
\\
&&p''\, z^3-\frac{z^2\,(1-3\,z^4)}{1-z^4}\, p'+\frac{3\, z^2\,(1-z^4)+\varphi^2}{(1-z^4)^2}\, p=0\ ,
\label{probeeom}
\eea
where the primes now denote derivatives with respect to $z$. As discussed above, the boundary asymptotics for the fields are

\begin{equation}
\varphi\rightarrow \frac{L^2\, \mu}{r_h}-\frac{4\pi^2\hat \rho\, L^6}{r_h^3}\, z^2+\cdots\qquad p\rightarrow \frac{L^3\, O_1}{ r_h}\, +\frac{L^3\, O_2}{r_h^3}\, z^2+\cdots
\label{asido}
\end{equation}
As  discussed in section 4.1, one must impose boundary conditions where $O_1=0$. 
Boundary conditions are implemented by solving the differential equations (\ref{probeeom}) by a standard shooting method.
{}From the solution one obtains $O_2=O_2(T)$.
Figure \ref{condTprobe} shows the curve corresponding to the condensate  $O_2\equiv \langle O\rangle$ as a function of temperature. 
The main new feature with respect to the finite $q$ case is that at $q\to\infty $ the condensate curve extends all the way down to $T=0$.
This confirms the tendency previously inferred from figures 3 and 4, where it is seen that, as $q$ increases, the condensed phase gets extended to lower temperatures.

Figure 7 is also in numerical agreement with figure 3, previously computed  including the back reaction. For comparison, the temperature in figure \ref{condTprobe} must be rescaled by 
$q^{1/3}$, owing to the rescaling of $\varphi$  by $1/q$, which produces a rescaling of
$\hat \rho/r_h^3$ in the asymptotic expansion (\ref{asido}). We recall that the temperature is obtained 
from the numerical solution by reading the value of $r_h$ from  the second term in (\ref{asido}), which will be thus rescaled by a factor $q^{1/3}$.
Therefore, at large $q$, $T_2 (q) = q^{1/3} T_{\rm probe},$ $T_{\rm probe}\approx 0.722$.
The critical temperature $T_2/q^{1/3}$ as a function of $q$ (obtained from the upper branch in fig. \ref{RomansTc}) is shown in figure 8.

\begin{figure}[h!]
\centering
\includegraphics[scale=.55]{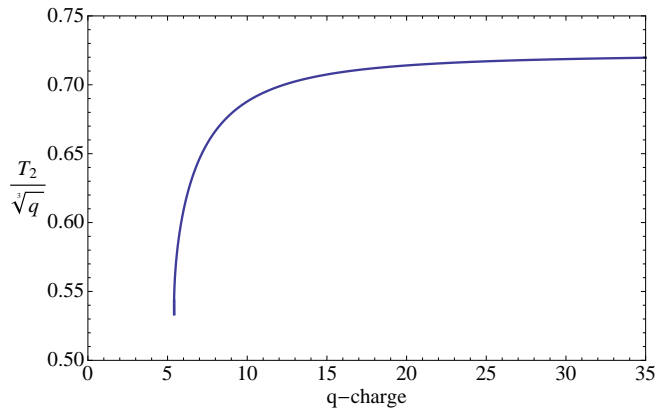} 
\caption{Critical temperature $T_2/q^{1/3}$ as a function of $q$.}
\label{proofprobe}
\end{figure} 

\begin{figure}[h!]
\centering
\includegraphics[scale=1]{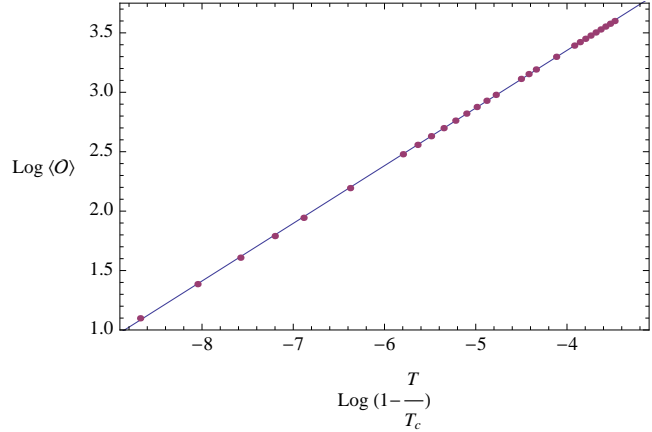} 
\caption{Fit of the logarithmic plot of  $\langle O\rangle $ vs. $T$ by a straight line of slope $\sim .49$.}
\label{criticalexp}
\end{figure}

The universality class of the transition can be inspected by computing the critical exponent $\beta $ in $\langle O\rangle \sim (T_c-T)^\beta $. 
Figure \ref{criticalexp} is a logarithmic  plot of $\langle O\rangle $ vs. $T$, which can be accurately fitted by 
a straight line.  We find the critical exponent  $\beta \approx 0.49$. 
Modulo numerical errors (coming mostly from the estimate of the critical temperature in the fit), this indicates that our phase transition has mean field critical exponent, 
as expected.\footnote{Non-mean field behavior can be accommodated in phenomenological models of holographic superconductors
by means of non-analytic terms \cite{Franco:2009if,Aprile:2009ai,Aprile:2010yb}. Such terms are not expected  in
{\it classical} Lagrangians originating from string/M theory compactifications, though they 
might effectively be induced by quantum corrections (see \cite{Aprile:2010yb} for a discussion).}

One interesting feature of the probe limit is that it pushes to zero the undetermined, possible unstable, low $T$ phase by replacing it with the hairy BH. In fact, this was our first hope, which is indeed realized in this large $q$ limit.

\subsection{Conductivity}

Many  universal features of superconductors just follow from the fact that these materials exhibit a spontaneous breakdown of $U(1)$ gauge invariance.
One of these features is vanishing electrical DC resistance.
Holographically, this  phenomenon occurs in a manner which is entirely  analogous to the field theory counterpart, namely
the $U(1)$ breaking turns on a new term in the Maxwell equations playing the role of a current --the London current.
As long as this term is non-vanishing, one expects infinite DC conductivity, as there is a finite current even for an infinitesimal electric field.
In the present case, the simplest way to see the emergence of this term is by considering a time-dependent perturbation of the form
\be
B_z=b_z(r)\ e^{i\omega t}+b_z^*(r)\ e^{-i\omega t}\ .
\ee
Turning on any other component like $B_x$ or $B_y$, will lead to a complicated system of coupled equations. This is a reflection of the fact
that the $A_{\mu\nu}$ background has broken isotropy and $z$ is a preferred direction.


We will compute the conductivities in the probe limit described  by the Lagrangian  (\ref{Lp}) in the fixed Schwarzschild-Anti de Sitter geometry
(where $X=1$). The conductivity, arising from the retarded current-current correlator, can be computed by studying fluctuations of the gauge field around the condensed solution discussed in the previous subsection. In the complete system, where gauge field and two-form backreact on the geometry, off-diagonal metric components in the Minkowski would be sourced. However, in the probe limit at hand, a consistent solution is in fact given by the same ansatz (\ref{aqq}), (\ref{app}) for $A^{\mu\nu}$, with the addition of a small $A^{0r}$ component, which is then given by
\be
A^{0r}=- \frac{q}{\sqrt{g}}\ \big( b_z(r)\ e^{i\omega t}+b_z^*(r)\ e^{-i\omega t}\big) a(r)\ .
\ee
{}For the remaining components we have $A_{0i}=A_{ri}=A_{zi}=0$, $i=x,y$.
All other components of (\ref{dosform})  are  satisfied identically, except for the $xy$ component, which gives (\ref{eqetan}) plus an $O(B_z^2)$ correction that we neglect.
The $z$ component of the Maxwell equation (\ref{maxi}) (particularized for Schwarzchild-AdS) then reads
\be
\label{bluct}
b_z''+\left(2A'-B'+{h'\over h}\right) b'_z+ {\omega^2 e^{2B-2A}\over h^2} \ b_z =      {q^2e^{2B-4A}\over  h }\  a^2\ b_z \ .
\ee
The r.h.s represents the holographic analog of the London current. 

Since we are interested in computing conductivities, the equation (\ref{bfluct}) has to be solved imposing causal boundary conditions at the horizon. This requires ingoing wave conditions, which set, close to the horizon

\begin{equation}
b_z\rightarrow (1-z)^{-i\frac{\tilde{\omega}}{4}}\, \sum_{k=0} b_k\,(1-z)^k\ ,
\end{equation}
where we have introduced $z=r_h/r$ as in the previous subsection and $\tilde \omega $ is determined from the equation (see below).
Then, the conductivity can be extracted from the asymptotic behavior of the gauge field fluctuation. 
As discussed in \cite{Horowitz:2008bn}, the asymptotic behavior of the gauge field fluctuation in five dimensions involves a logarithmic term which has to be  reabsorbed by adding a suitable counterterm. 
Generically we have

\begin{equation}
b_z\rightarrow b_z^{(0)}+\frac{b_z^{(1)}}{r_h^2}\, z^2-\frac{b_z^{(0)}\,\omega^2}{2\, r_H^2}\,z^2\, \log z\ .
\end{equation}
Then, following \cite{Horowitz:2008bn}, the conductivity is given by

\begin{equation}
\sigma=-i\frac{2\, b_z^{(1)}}{b_z^{(0)}\, \omega}+i\frac{\omega}{2}\ .
\end{equation}
Provided $b_z^{(1)}$ is not zero, the imaginary part of the conductivity has a pole, ${\rm Im}(\sigma ) \sim 1/\omega $. By standard relations of complex analysis,
this pole is associated with a delta function $\delta(\omega )$ in the real part of the conductivity (see \cite{Hartnoll:2008kx,Aprile:2009ai,Gubser:2008wv} for  discussions).
Note that the London term in (\ref{bluct}) is crucial for the emergence of the delta function. 
Without this term, the  $\omega\to 0$ limit of the equation has a constant solution with $b_z^{(1)}=0$.
In the presence of this  term, $b_z^{(1)}$  cannot vanish, leading to ${\rm Re}\ \sigma \sim\delta (\omega )$ and thus
DC superconductivity.

\begin{figure}[h!]
\centering
\includegraphics[scale=1]{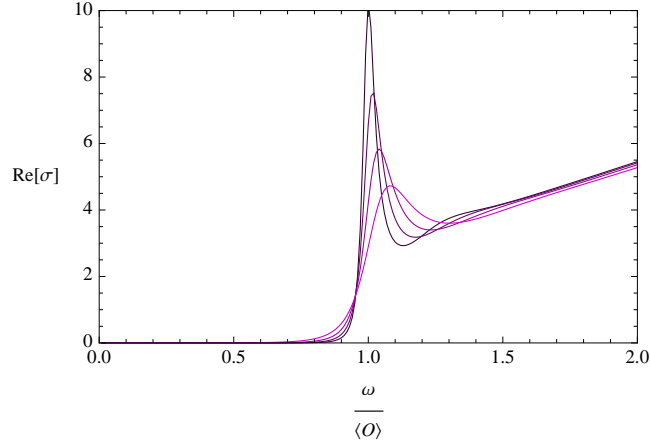} 
\caption{AC conductivity for $\frac{T}{T_c}=\{ 0.37,\ 0.41,\ 0.47,\ 0.54\} $
 (from bottom to top). }
\label{condomega}
\end{figure}

\begin{figure}[h!]
\centering
\includegraphics[scale=1]{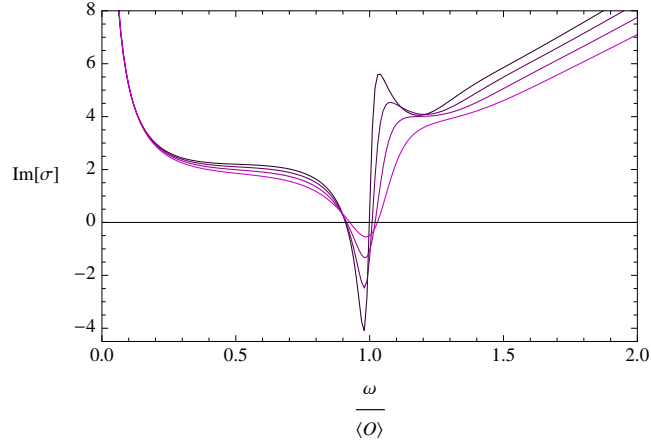} 
\caption{Imaginary part of the conductivity for $\frac{T}{T_c}=\{ 0.37,\ 0.41,\ 0.47,\ 0.54\}$
 (from top to bottom). }
\label{condomegaIM}
\end{figure}

Let us now explicitly compute the conductivity. 
 In this case, we can consider (\ref{bluct}) particularized to the Schwarzschild-Anti de Sitter background. We get
\begin{equation}
\label{bfluct}
b''-\frac{(1+3\,z^4)}{z\,(1-z^4)}\,b'+\frac{\tilde{\omega}^2-(1-z^4)p^2}{(1-z^4)^2}\, b=0\ .
\end{equation}
where $\tilde{\omega}=\frac{L^2}{r_h}=\frac{\omega}{\pi\, T}$.
Implementing the boundary conditions as described above, we now numerically compute the frequency-dependent conductivity.
The results are  shown in figures \ref{condomega} and \ref{condomegaIM}.
The imaginary part of the conductivity shown in figure \ref{condomegaIM} exhibits the $1/\omega $ behavior associated with a $\delta(\omega )$ behavior in the real part
(the delta function is not seen in figure \ref{condomega} due to numerical reasons).
%
{} 
There are basically three regimes for the frequency: 

\smallskip

\noindent a) $\omega< \langle O\rangle = O(\hat \rho^{1\over 3})$. Here the conductivity is strongly affected by the presence of the condensate 
and exhibits the expected gap in ${\rm Re}(\sigma )$ (related to the energy which is required to break the condensate).

\smallskip

\noindent b) $\langle O\rangle \ll \omega\ll T$.  Here the contribution from the condensate $p$ in  (\ref{bfluct}) can be neglected.
One is then  basically computing the conductivity of Schwarzchild-AdS black hole. 

\smallskip

\noindent c) $\omega \gg T $. Here the frequency is above all relevant scales and the conductivity approaches the conductivity of $AdS_5$, with a linearly increasing behavior. 

\smallskip


It is interesting to note the difference with  Schwarzchild-AdS$_4$ (discussed e.g. in \cite{Gubser:2008wv}), where the conductivity becomes constant at high frequencies, 
rather than linearly increasing (and the temperature does not play any role).
In contrast, in the case of Schwarzchild-AdS$_5$, the conductivity is linearly increasing at sufficiently large frequencies.
A detailed discussion can be found in \cite{Horowitz:2008bn}. A similar linear behavior also 
appeared in  \cite{Peeters:2009sr} in some $p$-wave four-dimensional holographic superconductor models based on probe D-branes.

%

\section{Comments on the dual field theory}

We have so far discussed a modification of $\mathcal{N}=4$ gauged supergravity in five dimensions in which a two-form field undergoes condensation as temperature is lowered. Now we would like to turn to its implications for the putative dual field theory. Even though our model cannot be obtained --or at least not in any obvious way-- as a consistent truncation of ten-dimensional IIB supergravity, 
the backgrounds we are considering asymptote to $AdS_5$. Therefore they are, on general grounds, subject to the general principles of AdS/CFT. In particular, we would like to determine the operator which is triggering the phase transition by its condensation. To that matter, let us consider the part of the action involving the 2-form,

\bea
I &\supset& -{1\over 16\pi G_N} \int \bigg[-\frac{1}{2}X^{-2}\,\ast A^{\a}_{(2)}\wg A^{\a}_{(2)}+\ \frac{L}{2}\eps_{\a\b}A^{\a}_{(2)}\wg dA^{\b}_{(2)}-\frac{q}{2}A^{\a}_{(2)}\wg A^{\a}_{(2)}\wg B_{(1)} \bigg]\ .
\eea
Dropping momentarily the interaction term with $B_{(1)}$, we have a first order action for a two-form potential with an effective mass given by $X^{-2}$. Note that, because the Lagrangian is first order, the parameter appearing in the Lagrangian is directly the mass of the field. In turn, the $X$ field has a potential given by

\begin{equation}
V=\frac{4}{L^2}\, \Big(X^2+2X^{-1}\Big)\ .
\end{equation}
The minimum of this potential --which the field must approach at large $r$-- is located at $X=1$. Therefore,  in the asymptotic region, the two-form behaves as a two-form field with mass $=1$, governed by a first order Lagrangian in $AdS_5$. We can then borrow the results in \cite{Arutyunov:1998xt}, where it was found that a mass $m$ two-form field in $AdS_5$ with first order Lagrangian couples to a boundary operator of dimension

\begin{equation}
\Delta=2+m\ .
\end{equation}
 Particularizing this expression to our case, we see that the $A_{(2)}$ fluctuation is dual to a dimension 3 operator which transforms as an antisymmetric tensor under the Lorentz group. This dual operator will live on the boundary strongly coupled CFT, which generically is expected to contain a gauge sector together with a matter sector. Since our dual operator must have two antisymmetric Lorentz indices, it must contain either two derivatives or else an insertion of a boundary gauge field strength. The first case would then require at least two scalars (or fermions), which would necessarily involve a dimension larger than 3. It is then natural  to guess that the dual operator contains $F_{\mu\nu}$. In order to have a non-vanishing trace, it is natural to conjecture that the dual operator is of the form $\mathcal{O}_{\mu\nu}\sim \Phi\, F_{\mu\nu}$, being $\Phi$ a certain scalar field of dimension 1 \footnote{Note that the inserted operator $\Phi$ has dimension 1, and thus corresponds to a free field. Despite the lack of explicit embedding into IIB and the lack of SUSY, it seems reasonable to expect that indeed 3 is the minimal dimension that  can be attained.}, such that  $\Delta_{\mathcal{O}}=3$. Besides, the bulk 2-form
potential  is charged with charge $q$ under the $U(1)$ field $B_{(1)}$, which suggests that the dual theory contains a global symmetry under which $\Phi$ has charge $q$, so that the quantum numbers of our proposed $\mathcal{O}_{\mu\nu}$ would match the expectations from the bulk physics. 

It is perhaps instructive to consider the case $q=1$ as test of our proposal. At $q=1$ our modified model becomes $\mathcal{N}=4$ $SU(2)\times U(1)$ Romans gauged SUGRA, which is a IIB consistent truncation. From the 10d point of view, the gauge field $B_{(1)}$ is turned on by angular momentum, and the solution can be thought as a stack of D3 branes in $\mathbb{R}^6$ spinning in a $U(1)\subset SO(6)$. This global symmetry is actually the R-symmetry of the theory, so that the background is dual to $\mathcal{N}=4$ SYM  at finite charge density in a $U(1)$ subgroup of the R-symmetry. The two-form is charged under this $U(1)_R$ subgroup, and naturally corresponds to $\Phi_3\, F_{\mu\nu}^+$, where $\Phi_3$ is a complex scalar of the $\mathcal{N}=4$ SYM theory so that the operator indeed has charge $q=1$ under the relevant $U(1)$ and dimension 3. We can understand this identification by slightly moving in the Coulomb branch of the theory higgsing the gauge group from $SU(N_c)$ to $SU(N_c-1)\times U(1)$. At large $N_c$ we can think of the system as a probe $D3$ brane moving in the background generated by the remaining $N_c-1\sim N_c$. The action for such a probe D3 is

\begin{equation}
S=-T_3\int e^{-\phi}\, \sqrt{{\rm det}\Big(P[g]+\mathcal{F}\Big)}+T_3\int \sum\, P[C_n]\wedge e^{\mathcal{F}}\ ,
\label{ddtres}
\end{equation}
where $\mathcal{F}=P[B]+2\pi l_s^2\,F$. Let us re-write the DBI determinant as

\begin{equation}
\sqrt{{\rm det}\Big(P[g]+\mathcal{F}\Big)} = \sqrt{{\rm det}P[g]}\, \sqrt{\Big( 1+P[g]^{-1}\,\mathcal{F}\Big)}\ .
\end{equation}
Then, expanding to second order the DBI we find a term with

\begin{equation}
S_{DBI}\supset \frac{T_3}{2}\int \mathcal{F}\wedge \star \mathcal{F}\ ,
\end{equation}
where the Hodge-star is taken with respect to the pull-back metric. In particular, this contains

\begin{equation}
S_{DBI}\supset T_3\, 2\pi l_s^2\int P[B]\wedge \star F\ .
\end{equation}
On the other hand, from the Wess-Zumino part of the D3-brane action (\ref{ddtres}), we find a term with

\begin{equation}
S_{WZ}\supset T_3\, 2\pi l_s^2\int P[C_2]\wedge F\ .
\end{equation}
Thus, altogether we have 

\begin{equation}
S_{D3}\supset T_3\, 2\pi\, l_s^2\, \int P[C_2]\wedge F+P[B]\wedge \star F\ .
\end{equation}
If $F$ is a self-dual two-form field, it follows that

\begin{equation}
S_{D3}\supset T_3\, 2\pi\, l_s^2\, \int \Big(P[C_2]+\, P[B]\Big)\wedge F^+\ .
\end{equation}
At this point it is useful to recall the truncation ansatz in  \cite{Lu:1999bw}, where  the starting point for the reduction on the 5-sphere $\sum_1^6 x_i^2=1$ is the following coordinate system

\begin{displaymath}
\begin{array}{l l l}
x_1 = \sin\xi\cos\tau\, ; & x_2 = \sin\xi\sin\tau\, ; & x_3 = \cos\xi\cos\alpha_1\, ;\\
x_4 = \cos\xi\sin\alpha_1\cos\alpha_2\, ; & x_5 = \cos\xi\sin\alpha_1\sin\alpha_2\cos\alpha_3\, ; & x_6 = \cos\xi\sin\alpha_1\sin\alpha_2\sin\alpha_3\, .
\\ \nonumber
\end{array}
\end{displaymath}

In terms of these angles, the 10-dimensional $B_2$, $C_2$ potentials are written in terms of $A_{(2)}$ as

\begin{equation}
B_2=-\sin\xi \sin \tau \, {\rm Re}\, A_{(2)}+\sin\xi\cos\tau\,  {\rm Im}\, A_{(2)}\, ;\quad C_2=-\sin\xi \cos \tau \, {\rm Re}\, A_{(2)}-\sin\xi\sin\tau\,  {\rm Im}\, A_{(2)}\ .
\end{equation}
Going back to the Cartesian coordinates we obtain

\begin{equation}
B_2 = - x_2 {\rm Re}A_{(2)}+x_1 {\rm Im}A_{(2)} \, ; \quad C_2 = - x_1 {\rm Re}A_{(2)}-x_2 {\rm Im}A_{(2)}\, .
\end{equation}
Therefore, we have

\begin{equation}
P[C_2+B_2]=  -(x_1+ x_2) {\rm Re}A_{(2)}+(x_1-x_2) {\rm Im}A_{(2)} \ .
\end{equation}
Let us now define  $\Phi=e^{i\frac{\pi}{4}} \, (x_1+i\, x_2)$. This field naturally corresponds to one of the three complexified scalars of $\mathcal{N}=4$ SYM, and it is then charged under a $U(1)$ subgroup inside the maximal torus of $SO(6)$. Conversely, we have

\begin{equation}
\Phi=e^{i\frac{\pi}{4}} \, (x_1+i\, x_2)=\sin\xi \, e^{i\frac{\pi}{4}}\, e^{i\tau}
\end{equation}
which explicitly shows  that the 
 $U(1)$ symmetry under which $\Phi $ has charge 1 corresponds to rotations in the $\tau$ direction from the ten-dimensional perspective. Furthermore,  the action of the probe brane can now be written as

\begin{equation}
S_{D3}\supset T_3\, 2\pi\, l_s^2\, \int \Big(P[C_2]+P[B]\Big)\, \wedge F^+ = T_3\, \sqrt{2}\pi\, l_s^2\, \int A_{(2)}\, \wedge \, \Phi\, F^+ +{\rm c.c.}
\end{equation}
This equation explicitly shows how $A_{(2)}$ couples to the proposed operator. Note that this operator is non-vanishing already at the abelian level, and thus there is no need to consider the non-abelian extension of the D3 brane action. 
Nonetheless, repeating the same exercise for the non-abelian case,  we obtain the straightforward extension of the above coupling,  ${\rm Tr}\Big(\Phi\, F^+\Big)$. 

It should be stressed that  it is actually the self-dual part of the gauge field strength what enters the dual operator. The reason is that $\mathcal{N}=4$ gauged SUGRA fixes the axiodilaton to zero. This bulk field  couples to the boundary operator $F_{\mu\nu}F^{\mu\nu}$. Since the axiodilaton is not turned on, it must be the self-dual part of $F$ the relevant one entering in the operator dual to the complex 2-form. Indeed, the proposed operator actually agrees with the results in \cite{Ferrara:1998ej} (see eq. (A.7)), in support of our conjecture.

It is interesting to compare with the picture of \cite{Gubser:2009qm}, where R-charged black holes undergoing a phase transition through the condensation of a chiral primary operator in IIB
theory were considered. Through AdS/CFT, these systems can be thought of as finite-temperature versions of D3 branes probing the tip of $CY_3$ cones which are $U(1)$ fibrations over a Sasaki-Einstein space. The $U(1)$ fiber corresponds to the R-symmetry of the theory, and by means of a chemical potential on such $U(1)_R$ charge, condensation is achieved.\footnote{We stress that this $U(1)_R$ is different from the one we turned on  in ${\cal N}=4$ $SU(2)\times U(1)$ supergravity. In the context of generalized STU black holes it would correspond to $q_1=q_2=q_3$; see section 6.} Interestingly, the operator condensing is also dimension 3. However, in this case corresponds to a chiral primary operator, and thus it has $R=2$. Indeed, it was argued to correspond to the superpotential of the theory. It is a subtle question whether this is the leading instability, since, depending on the theory under consideration, there might be R-charged chiral operators with lower R-charge. In our case, due to the vector-like nature of the condensing operator, the dual operator should contain, as discussed above, either two derivatives or an insertion of the field strength.\footnote{One might also wonder why an operator made out of two fermions with the suitable Dirac matrix does not do the job. The reason is that the combination $\bar{\Psi}\Gamma_{\mu\nu}\Psi$ would be neutral under the $U(1)_R$, and thus would require yet another scalar, thus making the dimension higher than 3.} The case with two derivatives would correspond to an operator of dimension at least 2. In order to have a gauge invariant operator with non-vanishing trace we would need at least two scalars, thus leading to a  dimension greater than 3. Thus, we see that the considered operator is the lowest dimensional one which can trigger the vector-like condensation.

\section{Search for 2-form condensation in $\mathcal{N}=8$ $SO(6)$ gauged Supergravity}

\subsection{ 2-forms in $\mathcal{N}=8$ $SO(6)$ gauged Supergravity}

Ungauged 5d SUGRA contains 42 scalars parameterizing the symmetric space $E_{6(6)}/USp(8)$. They can be compactly grouped in the 27-bein $V_{AB}^{ab}$, being $\{A,\,B\}$ $E_{6(6)}$ indices and $\{a,\, b\}$ $USp(8)$ indices. When turning to the gauged version, the natural $SO(6)$ gauged subgroup is embedded in a maximal $SL(6,\,\mathbb{R})\times SL(2,\,\mathbb{R})$ subgroup of $E_{6(6)}$. Since the fundamental representation of $E_{6(6)}$ breaks under $SL(6,\,\mathbb{R})\times SL(2,\,\mathbb{R})$ as $\mathbf{27}\rightarrow (\mathbf{\tilde{15}},\,\mathbf{1})\,+\,(\mathbf{6},\,\mathbf{2})$, the $A,\,B$ indices are split into $I=1\cdots 6$ and $\alpha=1,\,2$.

The gauged $\mathcal{N}=8$ SUGRA comprises both one-form and two-form gauge potentials. The two-form fields are in the $(\mathbf{6},\,\mathbf{2})$ and therefore have indices ${\cal A}_{\mu\nu}^{I\alpha}$. In turn, the one-form fields carry indices ${\cal B}_{\mu,\, IJ}$ in the $(\mathbf{\tilde{15}},\,\mathbf{1})$. From \cite{Gunaydin:1984qu, Gunaydin:1985cu} \footnote{Note that there is a factor of $1/8$ of difference in the conventions of these two references. Here we  follow the conventions of \cite{Gunaydin:1985cu}.} we can read off the relevant Lagrangian for the $\mathcal{N}=8$ two-forms ${\cal A}_{\mu\nu}^{I\alpha}$ coupled to the $SO(6)$ gauge field ${\cal B}_{\mu,\, IJ}$

\begin{equation}
\mathcal{L}\supset\,-\frac{1}{4}\,\epsilon^{\mu\nu\rho\sigma\tau}\,\epsilon_{\alpha\beta}\,\eta_{IJ}\, {\cal A}_{\mu\nu}^{I\alpha}\partial_{\rho}{\cal A}_{\sigma\tau}^{J\beta}+\frac{1}{2}\,\epsilon^{\mu\nu\rho\sigma\tau}\,\epsilon_{\alpha\beta}\,{\cal B}_{\mu,\, IJ}\, {\cal A}_{\nu\rho}^{I\alpha}\, {\cal A}_{\sigma\tau}^{J\beta}-\frac{1}{2}\,{\cal A}_{\mu\nu}^{I\alpha}\, {\cal A}^{\mu\nu,\, J\beta}\, M_{I\alpha,\, J\beta}
\end{equation}
where the mass matrix is given in terms of the scalar vierbein as

\begin{equation}
M_{I\alpha,\, J\beta}=V_{I\alpha}^{ab}\,V_{J\beta,\,ab}
\end{equation}
In the $SL(6,\,\mathbb{R})\times SL(2,\,\mathbb{R})$ subsector of the scalar manifold, the vielbein simplifies into \cite{Gunaydin:1985cu}

\begin{equation}
V_{I\alpha}^{ab}=\frac{1}{2\sqrt{2}}\, \Big(\Gamma_{K\gamma}\Big)^{ab}\, S^K_I\, \tilde{S}^{\gamma}_{\alpha}\, ,\qquad \Gamma_{K\sigma}=\Big(\Gamma_K,\, i\, \Gamma_K\,\Gamma_0\Big)\ ,\quad \sigma=1,\,2\ ,
\end{equation}
where $S\in SL(6,\,\mathbb{R})$ and $\tilde{S}\in SL(2,\,\mathbb{R})$. The matrices $\Gamma_i$ are the $SO(7)$ gamma matrices satisfying $\{\Gamma_i,\, \Gamma_j\}=2\,\delta_{ij}$ for $i,\, j=0\cdots 6$. Latin indices are raised and lowered with $\Omega^{ab}=-\Omega_{ab}=-i\, \Big(\Gamma_0\Big)^{ab}$. 

It is now a straightforward exercise to show that

\begin{equation}
M_{I\alpha,\, J\beta}=V_{I\alpha}^{ab}\,V_{J\beta,\,ab}=-M_{IJ}\, \tilde{M}_{\alpha\beta}
\end{equation}
being $M$ and $\tilde{M}$ the $SL(6,\, \mathbb{R})$ and $SL(2,\, \mathbb{R})$ metrics respectively. Explicitly

\begin{equation}
M_{IJ}=S_I^K\, S^K_J    \ , \qquad \tilde{M}_{\alpha\beta}=\tilde{S}^{\gamma}_{\alpha}\, \tilde{S}^{\gamma}_{\beta}\ .
\end{equation}

We will be interested in the scalar sector which is singlet under the $SL(2,\, \mathbb{R})$, and thus we can just take $\tilde{M}_{\alpha\beta}=\delta_{\alpha\beta}$. Furthermore, we would like to consider a  breaking of the gauge group down to $SO(2)\times SO(2)\times SO(2)$. This symmetry breaking pattern can be encoded in the scalar matrix $M$

\begin{equation}
\label{M}
M_{IJ}={\rm diag}\Big(X_1,\, X_1,\, X_2,\, X_2,\, X_3,\, X_3\Big)
\end{equation}
Since this is an $SL(6,\,\mathbb{R})$ matrix, its determinant must be one, which implies

\begin{equation}
X_1\, X_2\, X_3=1\ .
\end{equation}
Let us now consider the gauge field kinetic term. This can be written as

\begin{equation}
\mathcal{L}\supset-\frac{1}{2}\, F_{\mu\nu,\, ab}\, F^{\mu\nu,\,ab}=-\frac{1}{2}\, F_{\mu\nu,\, IJ}\, F^{\mu\nu}_{KL}\, V^{IJ,\, ab}\, V^{KL,\, cd}\, \Omega_{ac}\, \Omega_{bd}\ .
\end{equation}
Following  \cite{Gunaydin:1985cu}, the $(\mathbf{15},\,1)$ part of the scalar vielbein is written as

\begin{equation}
V^{I,\, ab}=\frac{1}{8}\, \Big(\Gamma_{KL}\Big)^{ab}\, U^{IJ}_{KL}\ ,\qquad U^{IJ}_{KL}=2\, S_{[K}^{\quad [I}\, S_{\quad L]}^{J]}\ .
\end{equation}
After some algebra, the kinetic term for the gauge fields becomes

\begin{equation}
\mathcal{L}\supset-\frac{1}{2}\, F_{\mu\nu,\, IJ}\, F^{\mu\nu}_{KL}M^{IL}\, M^{JK}\ .
\end{equation}
Taking into account our diagonal form for $M$ we have

\begin{equation}
\mathcal{L}\supset- \frac{2}{X_1^2}\,F_{\mu\nu,\, 12}\, F^{\mu\nu}_{12}- \frac{2}{X_2^2}\,F_{\mu\nu,\, 34}\, F^{\mu\nu}_{34}- \frac{2}{X_3^2}\,F_{\mu\nu,\, 56}\, F^{\mu\nu}_{56}\ .
\end{equation}
This exhibits in an explicit manner  the symmetry breaking pattern, which leaves an unbroken $U(1)^3$ gauge group whose three gauge potentials 
are $\{{\cal B}_{12},\, {\cal B}_{34},\,{\cal B}_{56}\}$. Defining 

\begin{equation}
{\cal B}_{\mu,\,12} = \frac{1}{2}\, B_{\mu}^1 \ ,\qquad {\cal B}_{\mu,\,34}= \frac{1}{2}\, B_{\mu}^2\ ,\qquad {\cal B}_{\mu,\,56}= \frac{1}{2}\, B_{\mu}^3\ ,
\end{equation}
the kinetic term for the gauge fields becomes
 
 \begin{equation}
\mathcal{L}\supset-\frac{1}{2}\sum\, \frac{1}{X_i^2}\, F_{\mu\nu}^i\, F^{\mu\nu\, i},\qquad F_{\mu\nu}^i=\partial_{[\mu}B_{\nu]}^i\ .
\end{equation}

Finally, let us consider the Lagrangian for the two-forms. We consider the following subsector:

\begin{equation}
 {\cal A}_{\mu\nu}^{2i-1\,\, 1}={\cal A}_{\mu\nu}^{2i\, \,2}
\ ,\qquad {\cal A}_{\mu\nu}^{2i-1\, \,2}=-{\cal A}_{\mu\nu}^{2i\, \,1}
\ ,\qquad i=1,\,2,\,3\ .
\end{equation}
It is convenient to define the complexified 2-forms

\begin{equation}
A_{(2)}^i={\cal A}^{2i-1\, \,1}+i\, {\cal A}^{2i-1\, \, 2}\ .
\end{equation}
Then, the Lagrangian  becomes

\begin{equation}
\mathcal{L}\supset \frac{i}{2}\Big\{ \, \epsilon^{\mu\nu\rho\sigma\tau}\, \bar{A}^i_{(2)\, \mu\nu}\partial_{\rho} A^i_{(2)\, \sigma\tau}- i\, \epsilon^{\mu\nu\rho\sigma\tau}\, B^i_{\mu}\, A^i_{(2)\,\nu\rho}\,\bar{A}^i_{(2)\, \sigma\tau}-2i\,\sqrt{g}\, X_i\, A^i_{(2)\,\mu\nu}\, \bar{A}_{(2)}^{i\,\mu\nu}\,\Big\}\ .
\end{equation}

\subsection{General STU black holes}

$\mathcal{N}=8$ gauged Supergravity is expected to arise  as a consistent truncation from ten-dimensional type IIB theory (even though this truncation has not been explicitly found yet).
Under the light of the ten-dimensional interpretation, the above symmetry breaking pattern stands for spinning branes in flat $\mathbb{R}^6$. The maximal torus in $SO(6)$ is $U(1)^3$, therefore there are at most three independent angular momenta. From the 5d perspective, the gauge fields $B^i$ are precisely associated with the $U(1)$ symmetries generated by three possible spins. In particular, electric charge under these gauge fields  uplifts to 10d as angular momenta in the corresponding Cartan plane of rotation.

The family of black hole solutions with three angular momenta arising from spinning D3 branes are precisely  the STU black holes of which a particular one-charge
case was given in section 2.
 The general solution for three different charges that is naturally embedded in  $\mathcal{N}=8$ $SO(6)$ gauged Supergravity is \cite{Behrndt:1998jd}

\bea
&&ds^2=-f\, H^{-2/3}\, dt^2+H^{1/3}\, f^{-1}\, dr^2+H^{1/3}\, \frac{r^2}{L^2}\, d\vec{x}^2 \ ,
\nn\\
\label{STUBH}
&&B_0^i=\frac{Q_i\, \sqrt{m}}{r_h^2+Q_i^2}-\frac{Q_i\, \sqrt{m}}{r^2+Q_i^2}\ ,\qquad X_i=H_i^{-1}\, H^{1/3}\ ,
\\
&&
f=\frac{r^2}{L^2}\, H-\frac{m}{r^2}\ ,\qquad H=H_1\,H_2\,H_3\ ,\qquad H_i=1+\frac{Q_i^2}{r^2}\ .\nn
\eea
%
For $Q_1=Q$, $Q_2=Q_3=0$ this solution becomes precisely the solution (\ref{atun}) discussed in previous sections, upon defining $X=X_1^{-1/2}$. In the following, we will be mostly interested in the case 
$Q_1=Q_2=Q_3=Q$ (with no loss of generality we can assume $Q\geq 0$). It is  useful to rescale $r\to r_h r$ so that the horizon is located at $r_h=1$ and define $Q=r_h\, \tilde{Q}$ and $m=r_h^4\,\tilde{m}$. Then, the horizon equation $f=0$ implies $\tilde{m}=(1+\tilde{Q}^2)^3$. In terms of the new variables the Hawking  temperature takes the form 
\begin{equation}
T=(2-\tilde{Q}^2)\,\,\frac{r_h \sqrt{1+\tilde{Q}^2}}{2\pi L^2}\ .
\end{equation}
On the other hand, the three electric field potentials become equal. For large $r$, they behave as

\begin{equation}
B_0\sim -\tilde{Q}\, \sqrt{1+\tilde{Q}^2}\, {r_h\over L}+\frac{\tilde{Q}\, (1+\tilde{Q}^2)^{3/2}\, r_h^3}{L\ r^2}+\cdots
\end{equation}
Thus, using the usual AdS/CFT prescription, we can read off the charge density 

\begin{equation}
\hat{\rho}=\frac{\tilde{Q}\, (1+\tilde{Q}^2)^{3/2}}{4\pi^2 L^3}\, r_h^3\ .
\end{equation}
Thus 
\begin{equation}
T=\frac{(2-\tilde{Q}^2)}{(2\pi)^{1/3}\, L\,\tilde{Q}^{1/3}}\, \hat{\rho}^{1/3}\ .
\label{TTrho}
\end{equation}
The temperature vanishes as $\tilde Q\to \sqrt{2}$. In terms of the original variables this corresponds
to $Q^2=2L\sqrt{m}/(3\sqrt{3})$.
{}  For  $\tilde Q>\sqrt{2}$ the geometry has a naked singularity and therefore no Hawking temperature.
 Thus $\tilde Q$ is restricted to be in the interval  $\tilde{Q}\in [0,\, \sqrt{2}]$. 
The maximal value of $\tilde{Q}$ corresponds to
\begin{equation}
r_h=\sqrt{\frac{2}{3}}\, \pi^{2/3}\, L\, \hat{\rho}^{1/3}\ .
\end{equation}
It is worth stressing that this value is not zero. Therefore,  the horizon  has a finite area in this  limit.

\vspace{.5cm}


\subsection{Condensation}

 $\mathcal{N}=8$ gauged SUGRA contains two-forms which could potentially condense, leading to superconducting transitions of the same nature
as in section 4, between R-charged black holes and black holes with 2-form hair. Thus it provides an interesting framework to study these type of transitions
in the more general context of a three-charge STU black hole.
Just as we did earlier, we shall consider a more general model where the 2-form has electric charge $q$ (with $q=1$ being the case of  $\mathcal{N}=8$ SUGRA)
and search for a zero mode in the bald black hole background with suitable properties. In this manner we shall determine the critical temperature
as a function of the electric charge of the 2-form field.

\medskip

In order to fix our conventions, it is useful to look at the scalar potential and compare it with the potential used in the derivation of  the STU black hole
\cite{Behrndt:1998jd}. As shown in \cite{Gunaydin:1985cu}, the scalar potential for $\mathcal{N}=8$ gauged SUGRA in the $SL(6,\, \mathbb{R})$ sector can be written as

\begin{equation}
V=\frac{1}{2}\, \Big(2\, W_{ab}^2-W_{abcd}^2\Big)\ ,
\end{equation}
where $W_{ab}=W^c_{acb}$ and 

\begin{equation}
W_{abcd}=\frac{1}{8}\epsilon^{\alpha\beta}\eta^{IJ}\, \Big(\Gamma_{L\sigma}\Big)^{he}\, \Big(\Gamma_{K\gamma}\Big)^{fg}\, S^K_I\, S^L_J\, \tilde{S}^{\gamma}_{\alpha}\, \tilde{S}^{\sigma}_{\beta}\ .
\end{equation}
After some algebra, one can verify that the potential can be compactly written in terms of $M$ as

\begin{equation}
V=\frac{1}{2}\, \Big(\,(\eta^{IJ}M_{IJ})^2-2\,\eta^{IJ}\,M_{JK}\,\eta^{KS}\,M_{SI}\,\Big) \ .
\end{equation}
Particularizing for the explicit expression for $M$ in (\ref{M}), we have

\begin{equation}
V=4\sum_{i=1,2,3}\frac{1}{X_i},\qquad X_1\, X_2\,X_3=1\, ,
\end{equation}
which coincides with the standard expression for the scalar potential for STU black holes (see \textit{e.g.} \cite{Behrndt:1998jd, Gubser:2009qt}), thus fixing our conventions.

\subsubsection{Case of one charge}

As a warm up, let us start by considering the case of a single charge $Q_1$, which corresponds to the $\mathcal{N}=4$ gauged SUGRA studied in previous sections. 
Choosing with no loss of generality the rotation plane along the $1-2$ coordinates, we have that the mass  for the two-form $A_{(2)}^1$ is $X_1$. The potential in turn becomes

\begin{equation}
V=4\Big(\frac{1}{X_1}+2\sqrt{X_1}\Big)\ ,
\end{equation}
which agrees with (\ref{Roman}) with the identification $X_1=X^{-2}$ ($X=X_2=X_3$). Using the  expression for the $X_i$ fields in the STU black hole (\ref{STUBH}) we have
$X=H^{1/3}$. One also checks that the part of the Lagrangian containing the 2-form agrees with (\ref{Roman}).
 In this case the gauge sector only comprises the gauge potential $B^1$, whose kinetic term reads

\begin{equation}
-\frac{1}{2}\, X^4\, F_{\mu\nu}\, F^{\mu\nu}\ ,
\end{equation}
in agreement with (\ref{Roman}).

\subsubsection{Case of three equal charges}

\begin{figure}[h!]
\centering
\includegraphics[scale=1]{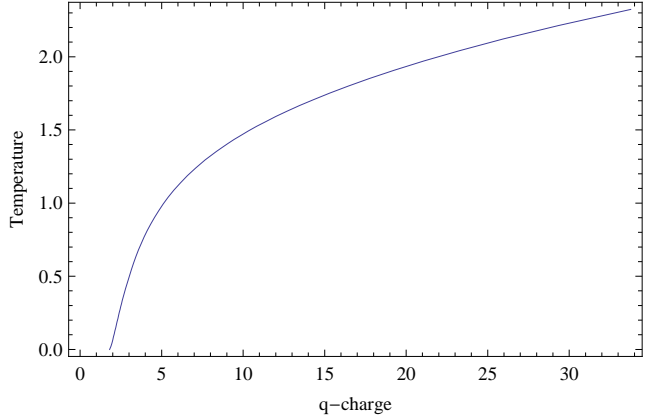} 
\caption{Critical temperature as a function of the 2-form $U(1)$ charge $q$.}
\label{Q1Q2Q3p}
\end{figure}

Let us now turn to the case of interest $Q_1=Q_2=Q_3=Q$. In this case the condition $X_1X_2X_3=1$ implies that $X_1=X_2=X_3=1$. 
As in previous sections, it is  instructive to consider
a more general Lagrangian where the 2-form has electric charge $q$. The three two-forms $A_{(2)}^i$ have the same Lagrangian, 
\begin{equation}
\mathcal{L}\supset \frac{i}{2}\Big\{ \, \epsilon^{\mu\nu\rho\sigma\tau}\, \bar{A}_{(2)\, \mu\nu}\partial_{\rho} A_{(2)\, \sigma\tau}- i q\, \epsilon^{\mu\nu\rho\sigma\tau}\, B_{\mu}\, A_{(2)\,\nu\rho}\,\bar{A}_{(2)\, \sigma\tau}-2i\,\sqrt{g}\, \, A_{(2)\,\mu\nu}\, \bar{A}_{(2)}^{\mu\nu}\,\Big\}
\end{equation}
We turn on, say $A_{(2)}^3$, and adopt the same ansatz for the 2-form components
as in section 3.1. We find the equation

\be\label{eqeta}
a''\ +\ \Big(\frac{h'}{h}+\frac{b'}{2b}-B'\Big)a'\ +\ {q^2\over L^2} e^{2B-2A}\frac{\Phi^2}{h^2}a\ -\ \frac{1}{L^2 h}e^{2B}\, a\ =0\ .
\ee
Defining again $Q=r_h\, \tilde{Q}$, $\bar{m}=r_h^4\, \tilde{m}$, $z=r_h/r$, and $p(z)= z\, a(z)$, the relevant equation is now
\be
p''(z)+F(z) p'(z)+G(z)p(z)=0\ ,
\label{tym}
\ee
with
\begin{eqnarray}
\label{eqpQ1Q2Q3}
F &=&  - \frac{ \Big( 1+2\,\tilde{Q}^2\, z^2+3\,(1+\tilde{Q}^2)^3\, z^4-\tilde{Q}^2\,z^6\,(1+3\,\tilde{Q}^2+5\,\tilde{Q}^4+\tilde{Q}^6)-\tilde{Q}^8\, z^8\Big)}
{z(1-z^2)\, (1+\tilde{Q}^2\, z^2)\, (1+(1+3\tilde{Q}^2)\, z^2-\tilde{Q}^6\, z^4)}\ ,
 \nn\\
G &=&  \frac{\tilde f}{(1-z^2)\,(1+\tilde{Q}^2\,z^2)\,(1+(1+3\tilde{Q}^2)\,z^2-\tilde{Q}^6\,z^4)^2}\ ,
\end{eqnarray}
and
\begin{eqnarray}
\tilde f  &= &
z^2 \left(z^4 \tilde Q^6-\left(3 \tilde Q^2+1\right) z^2-1\right) \left(\tilde Q^2
   \left(\left(\tilde Q^4 z^4+\left(\tilde Q^4+5 \tilde Q^2+3\right) z^2-3 \tilde Q^2-8\right)
   \tilde Q^2+z^2-9\right)-3\right)
\nn\\
&+& q^2 \tilde Q^2 \left(1+\tilde Q^2\right) \left(1-z^2\right)
   \left(1+ z^2 \tilde Q^2\right)^2\ .
\end{eqnarray}
%
The large $r$ (small $z$) asymptotic of the two-form is controlled by the mass parameter, and it is 
the same as in the case studied in section 4,
\begin{equation}
p(z)\rightarrow r_h O_1+O_2\ \frac{z^2}{r_h}+\cdots
\end{equation}
Thus, the normalizable zero-mode must satisfy  $p(z=0)=0$. In order to search for such a zero mode, we can numerically solve (\ref{eqpQ1Q2Q3}) as a function of $q$ and $\tilde Q\in [0,\, \sqrt{2}]$
and determine $\tilde Q=\tilde Q(q)$ from the condition $p(z=0)=0$. 
Using the formula (\ref{TTrho}) for the temperature we then find $T=T(q)$. The results are shown in figure \ref{Q1Q2Q3p}.

The figure shows that  there are hairy black holes provided $q>q_{\rm min}\approx 1.8$.
This excludes the case of  $\mathcal{N}=8$ $SO(6)$ gauged Supergravity where $q=1$.
In other words, the charge of the 2-form fields of $\mathcal{N}=8$ $SO(6)$ SUGRA is not large enough to drive to an instability.
The modified theory with arbitrary $U(1)$ charge $q$ has critical temperatures ranging from 0 to infinity.

{}For the minimal charge $q_{\rm min}\approx 1.8$  the   critical temperature vanishes.
(it corresponds to the extremal case $\tilde{Q}\to \sqrt{2}$).
One thus find that the theory with $q=q_{\rm min}$ has a normalizable zero-mode at zero temperature. 
Therefore, the theory has a quantum critical point.
It would be  interesting to study its properties (for $s$-wave holographic superconductors, studies of quantum critical points have recently appeared in \cite{Faulkner:2010gj}).

\subsection{Dual field theory operator}

We have seen that the theory described by the generalized Lagrangian inspired in $\mathcal{N}=8$ gauged SUGRA, where the 2-form field has electric charge $q$,
 undergoes a superconducting phase transition provided $ q\gtrsim 1.8$. 
 We now turn to a holographic interpretation along the lines of the $\mathcal{N}=4$ case discussed in  section 5. 

In the three-charge case the mass of the 2-form becomes a constant equal to one. Following \cite{Arutyunov:1998xt}, we conclude that it corresponds to a dimension three operator transforming as an antisymmetric Lorentz tensor. 

The natural operators which are dual to $A_{(2)}^i$ are
\begin{equation}
O_{\mu\nu}^i = {\rm Tr}\big[ \Phi_i\, F^+_{\mu\nu} \big]
\end{equation}
They all  have the same charge under the diagonal $U(1)$ subgroup in $U(1)^3\in SO(6)$. 
In $\mathcal{N}=4$ SYM this operator has $R$-charge equal to one.
 In our modified theory it must have  R-charge $q$ in order to match  quantum numbers of the two-form fluctuation.
Like in the $\mathcal{N}=4$ case discussed in  section 5, this identification can be motivated by considering the
non-abelian theory Born-Infeld theory with chemical potentials representing the rotating D3 branes.

\section{Concluding remarks}

In this paper we have investigated holographic $p$-wave superconductors emerging from 
the condensation of bulk 2-form fields in
models which are a slight modification of ${\cal N}=4,8$ five-dimensional gauged 
supergravities, where the two-forms are promoted
to two-forms of arbitrary $U(1) $ charge $q$. 
The gauged supergravity setup is particularly attractive given the explicit knowledge of  the dual field theory
given in terms of Super Yang-Mills theory with chemical potentials.
We found that condensation requires, at least for some specific ans\" atze, a minimal 
value of the charge $q$ which is, unfortunately,
above the value of the $U(1)$  charge  in supergravity. 
The model with general charge $q$ describes an alternative holographic realization of $p$-wave superconductivity
that complements previous approaches \cite{Gubser:2008wv,Roberts:2008ns,Ammon:2009fe,Peeters:2009sr,Basu:2009vv,Ammon:2009xh}.


\medskip

There are a number of very interesting open problems:

\begin{itemize}

\item One can similarly consider ${\cal N}=8$ $SO(8)$ four-dimensional gauged supergravity
as a framework for the study of $p$-wave superconducting models in three dimensions
based on condensation of charged vector fields.
It would be interesting to see if also in this case condensation requires  minimal 
charges which are above the gauged supergravity values.

\item One may also study condensation of charged vector fields in the context of 
 the ${\cal N}=4$ and ${\cal N}=8$ five-dimensional gauged supergravities
discussed in this paper.
In particular, in the ${\cal N}=4$  case one could consider  STU black holes charged with respect to the $U(1)\subset SU(2)$
and look for condensation of the $W$ vector bosons.

\item It would be interesting to elucidate more detailed condensed matter aspects of the
present models. In particular, to clear up the anisotropic properties of the conductivity.

\end{itemize}

\section*{Acknowledgements}

We would like to thank Oren Bergman, Diederik Roest and Oscar Varela for discussions. We also thank an anonymous referee for correcting an important point in an earlier version of this paper.
F.A. is supported by a MEC FPU Grant No.AP2008-04553. D.R-G. is supported by the Israel Science Foundation under grant no. 392/09. He also acknowledges support from the Spanish Ministry of Science through the research grant no. FPA2009-07122 and Spanish Consolider-Ingenio 2010 Programme CPAN (CSD2007-00042).
J.R. acknowledges support by MCYT Research
Grant No.  FPA 2007-66665 and Generalitat de Catalunya under project 2009SGR502.




\end{document}